\documentclass[structabstract]{aa}  
\usepackage{graphicx}
\usepackage{txfonts}
 \usepackage{amssymb}
  \usepackage{natbib}
  \usepackage{lscape}
  \bibpunct{(}{)}{;}{a}{,}{,}
  \usepackage{times}

  \usepackage{longtable}
\usepackage{longtable,lscape}
 \def\numwtt{65}
 \def\numctt{21}
 \def\numcl0{thre}
 \def\numirexc{24}
 \def\numnoclass{five}
 \def\numtarg{94}
 \def\numcomflames{68}
 \def\flamesmem{63}
 \def\flamesnomem{three}
 \def\flamesuncert{two}

 \def\numxdet{87}
\def\numnoflames{26}
\def\membdef{88}
\def\nomembdef{six}
\def\nkstars{42}
\def\nmstars{32}

\begin{document}
%
   \title{Spectral classification and HR diagram of pre-main sequence stars in 
NGC6530\thanks{Based on observations collected at the European Southern Observatory, Paranal,
  under program ID 077.C-0073B}}

   \subtitle{}

   \author{L. Prisinzano
          \inst{1}
          \and
          G. Micela\inst{1}
	  \and
	  S. Sciortino\inst{1}
	\and 
	 L. Affer\inst{1}
	\and
	F. Damiani\inst{1}
          }

   \institute{INAF - Osservatorio Astronomico di Palermo, Piazza del Parlamento
 1, 90134 Palermo\\
   \email{loredana@astropa.inaf.it}}
   \date{Received XXX; accepted XXX}

%
  \abstract
   {Mechanisms involved in the star formation process and in particular the   duration 
of the different phases of the cloud contraction are not yet fully understood.
Photometric data alone suggest that objects coexist in the young cluster NGC\,6530  with ages
from $\sim$1\,Myr up to 10\,Myrs.}
   {We want to derive accurate stellar parameters and, in particular, stellar ages 
to be able to constrain a possible age spread in the star-forming region NGC6530.}
   {We used low-resolution spectra taken with VIMOS@VLT and literature spectra of standard stars
 to derive spectral types of a subsample of 94 candidate members of this cluster.}
   {We assign  spectral types to 86 of the 88 confirmed cluster members and derive individual
reddenings. 
Our data are better fitted by the  anomalous
reddening law with R$_{\rm V}$=5. We confirm the presence of strong differential reddening in this region.
We derive fundamental stellar parameters, such as 
effective temperatures, photospheric  colors, luminosities, masses, and ages
 for 78 members, while for the remaining 8 YSOs we cannot determine the interstellar absorption,
since they are likely accretors, and their
 V-I colors are bluer than their intrinsic colors.}
   {The cluster members studied in this work have masses between 0.4 and 4\,M$_\odot$
and ages between 1-2\,Myrs and 6-7\,Myrs.
 We find that the SE region is the most recent site of star formation, while the older YSOs are loosely
clustered in the N and W regions. The presence of two distint generations of YSOs with different
spatial distribution allows us to conclude that in this region there is an age spread of $~$6-7\,Myrs. This is 
consistent with the scenario of sequential star formation suggested in literature.}

\authorrunning{L. Prisinzano et al.}
\titlerunning{HR diagram of PMS stars in NGC6530} 
\maketitle
%

\section{Introduction}
The study of very young open clusters is crucial  to understanding the process 
  of star formation within molecular clouds, since by measuring their age dispersion, 
it is possible to reconstruct the star formation history and test different star formation theories. A still controversial issue
  concerns the predictions of the duration  of different phases of star 
  formation. According to early models, clouds were considered in virial
 equilibrium,
which makes them evolve on long time scales
($\gtrsim 10^8$ years) \citep[e.g.][]{solo79,pall00,tan06,huff07}. In contrast, 
recent observations obtained with the {\it Herschel}
observatory reveal very elongated, filamentary structures \citep[e.g.][]{moli10},  
suggesting clouds affected
by supersonic turbulence \citep{ball07} that are expected to form stars 
on short time scales ($\sim 10^7$ years) 
\citep[e.g.][]{elme00,hart01b,elme07}.

Reliable age determinations in star-forming regions are thus crucial to deriving the
age spread and establish the duration of the star formation process. This allows us 
to constrain star formation theory and also to understand the evolution of the circumstellar material
around young stellar objects (YSOs), which are strictly related to planet formation.

Stellar ages can be derived by different methods, as discussed in 
\citet{sode10}, but the physical complexity of YSOs and of their environment
imply very large uncertainties on the stellar parameters from which ages
are inferred \citep[see][for a detailed review]{prei12}.
The most widely used method of estimating the ages of young stars is to derive
effective temperatures and luminosities from observational data and compare
them in the  Hertzsprung-Russell diagram (HRD) with those predicted by 
theoretical pre-main sequence (PMS) models. 
Uncertainties in the stellar parameters of SFRs, 
if not appropriately considered, can be underestimated  leading to the
interpretation the observed scatter in the HRD as an intrinsic age spread
rather than as un upper limit to the true age spread \citep{prei12}.

One of the most important contributions on the stellar parameter uncertainties comes
 from the presence of different amount of interstellar and circumstellar extinction around YSOs. 
This yields
highly unreliable derivations of effective temperature  from photometric colors
and therefore requires the acquisition of low-resolution spectra to derive spectral types to
infer accurate effective temperatures. An appropriate reddening law to derive the absorption and
therefore the luminosity, and a suitable treatment of other sources of uncertainties,
such as excess emission in the blue and/or in the IR spectrum, binarity, and photometric variability
are also crucial for deriving accurate stellar parameters, hence reliable stellar ages.


  A good target for studying all these topics is the very young and rich 
   open cluster NGC\,6530 (associated with the giant molecular cloud M8 
   or Lagoon Nebula),
     located about 1250 pc from the Sun. 
   A peculiarity of this region is the presence of
   the Hourglass nebula, a compact  H II region, produced by the 
   extremely young  O7.5 V   star Herschel 36.
   
   Several investigations have been
   devoted  to this cluster since the work of Walker (1957). The main
   challenge when studying star-forming regions comes from most of the
   cluster stars still being in the PMS phase and, due to the age spread and/or 
   observational uncertainties, are located in a broad  region
   of the color-magnitude diagram (CMD). 
   In the case of NGC\,6530, only stars hotter than about A0 spectral types
 are located on the cluster main sequence, 
   while most of the cluster members show a wide spread in the CMD.
    \citet{sung00}
   gives a list of 37 PMS candidates of NGC\,6530, based on $H_\alpha$ and 
   {\em UBVRI} photometry,
   down to $V\sim17$. They also derived individual reddening values
   for 30 O and B type stars,
   finding evidence of differential reddening with $E(B-V)$ in the range 
   [0.25,0.5] mag  and
   a mean value of 0.35.
   More recently, a long list of candidate members (884 sources) 
   has been obtained    using a deep Chandra ACIS-I
   X-ray observation \citep{dami04}.  Optical and IR (2MASS) 
   properties of these X-ray sources
   have been analyzed by us in  \citet{pris05} and in \citet{dami06}. 
    Using  $BVI$ images taken with 
   the WFI@ESO/2.2\,m
   telescope, we obtained a photometric catalog that reaches down to $V\simeq23$,
   from which we derived a revised
   cluster distance of $\simeq 1250$ pc. This value
   has been confirmed by \citet{aria06} 
   who derived the cluster
   distance using the spectral energy distribution, from  
   optical and IR photometry, for a group of stars around Herschel 36.
   Adopting  this distance value, the \citet{sung00} mean reddening and theoretical
    models,    we estimated ages and masses from photometry and, within the
large uncertanties derived by the assumption of a mean reddening, we found indications
of a sequential star formation. 

Few spectroscopic studies are available for the low-mass population of this cluster.
In particular,  332 PMS candidates with $V \le 18$ mag, chosen based on
   their position in the color-magnitude diagrams, have been studied using
   high-resolution
   spectroscopic observations (R=19\,300) taken with GIRAFFE-FLAMES@VLT, with the
   setup around the lithium line, and  around the $H_\alpha$  line for a 
    subsample of them \citep[][hereafter PDM2007]{pris07}. Most
   of them (237) are confirmed cluster members based on their radial velocities and on
   the presence of a strong lithium line.
   However, most of the derived properties
    of the cluster (age spread, disk frequency, IMF, and
    history of star formation) are based on temperatures derived by the B-V and V-I colors, 
assuming a mean reddening, so they are affected by very large uncertainties.

Intermediate-resolution spectra in the range 3850--7000\,$\AA$ 
of 39 YSOs in NGC\,6530 with H$\alpha$ emission and NIR excess
were used by \citet{aria07} (hereafter ABM) 
to derive effective temperatures, luminosities, masses, and ages,
by means of NIR photometry. Although the sample of studied members is relatively small, their results
 also suggest a sequential process in the star formation. 

  In this paper we present the spectral type classification
of a selected sample of candidate members of NGC\,6530 with a large number 
of relatively {\it clean} objects with X-ray emission but without evidence of NIR excesses
 and thus likely weak T Tauri stars (WTTS).  In addition, we present results for 
 YSOs with evidence of NIR excesses and for 
 candidate YSOs
 selected  from their position in the CMD.
The aim of this work  is to 
  determine reliable and accurate  effective 
   temperatures and luminosities to be placed in the H-R diagram.
In addition we want to investigate if a 
    real age spread is present in this cluster and if the
    evidence of a sequential and triggered star formation from N to S,
  as found   since the work of \citet{lada76}, can be confirmed with our data.

\section{Observations}
\subsection{Target selection}
We observed a total of \numtarg\,  spectroscopic targets, selected among candidate members
with $12.5<V<19.0$ in the PMS region of NGC\,6530  by using the \citet{pris05} 
optical catalog obtained from images taken with
the WFI camera of the ESO 2.2 m Telescope. 
 Most of our targets (\numwtt) do not show strong IR excesses and  
  were chosen for being likely {\it clean} YSOs  
not affected by severe veiling effects, for which the 
derived stellar parameters should be affected by relatively 
small uncertainties. In addition, we selected \numirexc\, candidate YSOs with 
evidence of NIR excesses in the JHK color-color diagram and \numnoclass\, further
targets selected because they fall  in the PMS region of the V vs. V-I diagram.

Among our targets, \numxdet\, have an X-ray counterpart in \citet{dami04}, while
\numcomflames\, are in common with the sample of VLT/FLAMES spectra 
published in PDM2007.
\addtocounter{table}{1} 
 In Table\,\ref{targettable}
 we list coordinates, literature identification number, and photometric data
of our targets, and in Fig.\,\ref{vivtarget} (top panel) we show
the V vs. V-I diagram where candidates without and with NIR excesses and 
selected only photometrically are highlighted with different symbols.

\subsection{Observations and data reduction}
Spectroscopic observations of the NGC\,6530 cluster were performed
on May 25, 2006 and on June 17, 2006 within the ESO program 077.C-0073B,
 with the VIMOS instrument mounted at the Nasmyth focus B of
the ESO-VLT UT3 telescope.

We used the multi-object spectroscopy (MOS) and
high-resolution (HR) modes, with the Orange setup 
(spectral resolution $R\sim 2150$, with a slit width of  1 arcsec),
covering a spectral range approximatevely between 5200 and 7600 $\AA$. 
Slit definition and positioning were performed with 
the VIMOS Mask Preparation Software (VMMPS).
Target positions were taken from  the source list found in
the VIMOS pre-imaging.
Our targets were observed with three exposures of 2260\,sec. Further details
are given on the log of observations given in Table\,\ref{logbook}.

Data were reduced with the VIMOS pipeline and the ESO Recipe Execution Tool (EsoRex).
For the analysis of this work  we considered the reduced spectra of detected objects.
For the spectra obtained on
 May 29, calibration lamps were acquired after seven and ten hours from the
 first and second observations, respectively. As a consequence, the telescope rotator position
was very different from  at the time of the science spectra acquisition.
Since the relative slit position with respect to the grism changes with the rotator position,
these spectra are affected by a distortion that caused a 
  variable shift along the dispersion direction up to 12\,$\AA$ in the 
 wavelength calibration.
For this reason, our spectra were not summed to improve the S/N but 
were analyzed separately, and appropriate local wavelength corrections have been found,
when required, for the analysis presented in this work.


\begin{table}
\caption{Log of the VIMOS observations}             
\label{logbook}      
\centering                          
\begin{tabular}{c c c c}        
\hline\hline                 
ID$_{\rm OB}$ & Date  & UT & Seeing (") \\    
\hline                        
237861 &2006-05-29 &04:22:20.622 &1.20\\
237861 &2006-05-29 &04:22:28.124 &1.20\\
237861 &2006-05-29 &04:22:28.125 &1.20 \\
237861 &2006-05-29 &04:22:20.621 &1.20\\
237870 &2006-05-29 &07:20:32.721 &1.64\\
237870 &2006-05-29 &07:20:39.428 &1.64\\
237870 &2006-05-29 &07:20:39.429 &1.64 \\
237870 &2006-05-29 &07:20:32.720 &1.64\\
237870 &2006-06-17 &03:35:39.135 &1.04\\
237870 &2006-06-17 &03:35:35.610 &1.04\\
237870 &2006-06-17 &03:35:35.609 &1.04\\
237870 &2006-06-17 &03:35:39.136 &1.04\\
\hline                                   
\end{tabular}
\end{table}

\section{Results}
\subsection{Identification of  cluster members}
\begin{figure}
   \centering
   \includegraphics[width=9cm]{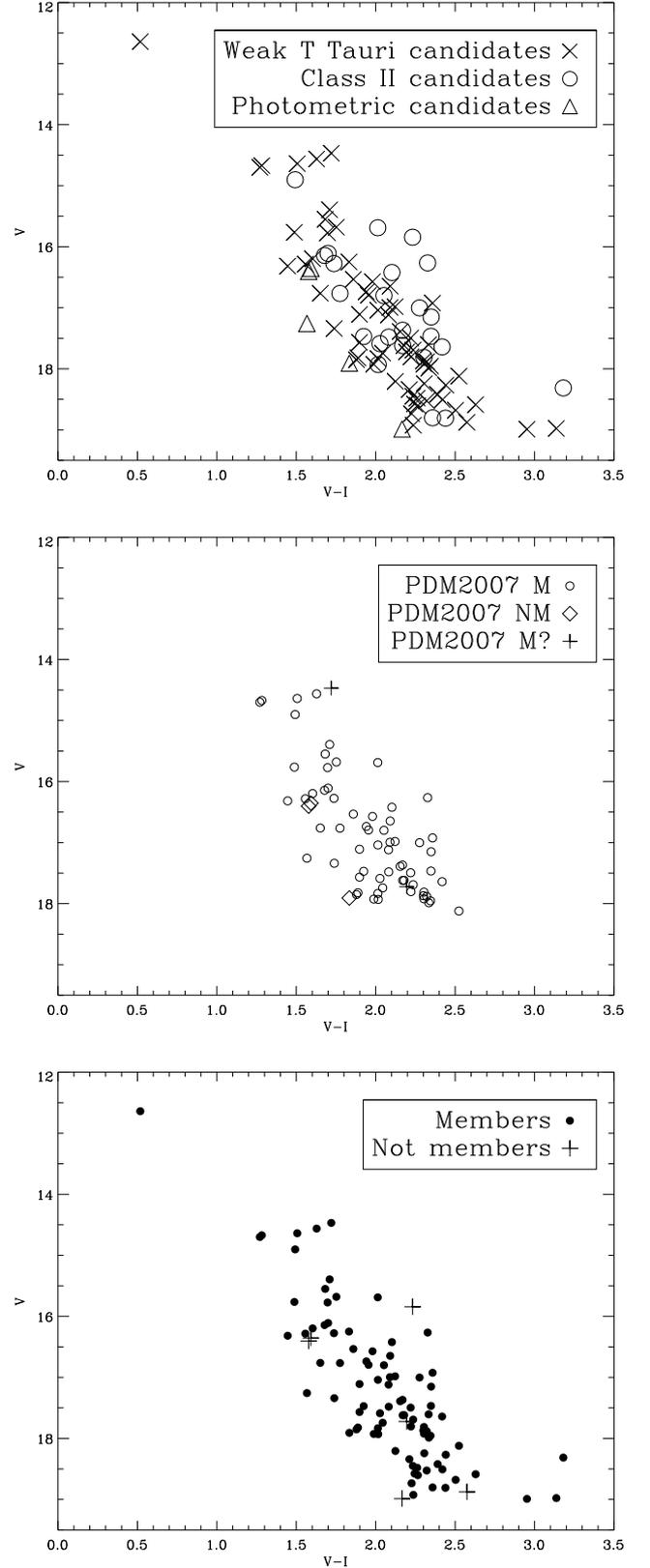}
      \caption{V vs. V-I diagram of our targets. Top panel shows the \numwtt\, targets
classified as candidate weak T Tauri stars, since they were detected in the X-rays,
the \numctt\, targets with IR excesses and the \numnoclass\, photometric candidates,
selected for the VIMOS observations. Middle panel shows the CMD for the \numcomflames\ targets,
also observed with FLAMES were `M' indicates the \flamesmem\, confirmed cluster members,
`NM' indicates the \flamesnomem\, nonmembers and `M?' indicates the \flamesuncert\, 
uncertain members. Bottom panel shows the CMD for the \membdef\, targets 
classified here as cluster
members  and the \nomembdef\, classified as nonmembers.}
         \label{vivtarget}
   \end{figure}

For the \numcomflames\, objects observed with both instruments 
\citep[VIMOS, this work and FLAMES,][]{pris07}, the
cluster membership  has been established 
by using the results from the FLAMES  high-resolution spectra
obtained by \citet{pris07},   based on radial velocities and on
the equivalent width of the Li I absorption line at 6707.8\,$\AA$.
Figure\,\ref{vivtarget} (middle panel) shows the V vs. V-I diagram of the VIMOS targets
with a counterpart in \citet{pris07}, where \flamesmem\,of them
 were classified as confirmed cluster members (`M'),
\flamesnomem\,   as cluster nonmembers (`NM') and \flamesuncert\, with uncertain membership (`M?').

In this work, we consider confirmed cluster members the  \flamesmem\, objects labeled in PDM2007 with `M' and 
the  star ID=24 labeled with `M?' for its discrepant radial velocity. Among the confirmed members
 we include 
the star with ID=19 that was classified in PDM2007 as a nonmember since this star does not show any X-ray detection,
Li line, or H$\alpha$ broadening in the FLAMES spectrum; neverthless, the VIMOS spectrum shows an evident 
Li absorption  and a broad H$\alpha$ in emission with FWZI $\sim 17\,\AA$. After careful analysis of the
FLAMES spectrum we noted that  the whole  spectrum shows very few lines, and we suspect there is a problem in the fiber assignment for this star. 
Thus we classify the star with ID=19 here as a confirmed
cluster member.

For those spectra that we only observed spectroscopically with VIMOS,   
 we established the membership status by 
considering the presence of the X-ray detection and by inspecting
the presence of the Li {\tiny I} absorption line at 6707.8\,$\AA$ 
and the shape of the H$\alpha$
line. Unfortunately, we cannot measure the equivalent width 
of  the Li {\tiny I}, since the spectral resolution of our spectra is too low to resolve 
this line, which is blended with the nearby iron lines   
\citep[Fe {\tiny I} 6705, 6710\,$\AA$,][]{covi97}.
However, since the Li {\tiny I} line is expected to be very strong for late type YSOs,
as are many of our targets, we used this line to assert the membership
for the \numnoflames\, objects for which we do not have  high-resolution spectra.

For YSOs, emission lines and, in particular, the
 H$\alpha$   line, are good indicators of youth,  but in the case of NGC\,6530,  
emission lines cannot be used as membership indicators, since the spectra may suffer from
H$\alpha$ line contamination from the surrounding Lagoon nebula. Owing to its strong spatial
variability, the subtraction of the nebula contribution is not reliable, and thus 
we did not attempt to improve the sky subtraction  obtained with the  
recipes of the VIMOS pipeline. Neverthless,  
the  H$\alpha$ line from the Nebula is expected to be  narrow, while that from YSOs with
accretion  is expected to be widened at the continuum level thanks to the gas motion
in the circumstellar regions. Even if it is very hard to accurately measure the full width
at zero intensity (FWZI) in low-resolution spectra, we inspected the base of the H$\alpha$ line,
so we considered cluster members those with a broad  line, i.e., those with FWZI $ \ge 10\,\AA$. 

In conclusion, we consider nonmembers to be the following objects:
stars ID=15 and 70, already classified as `NM' in PDM2007 for the absence of the Li line,
the lack of X-ray detection, and the H$\alpha$ in absorption;
stars  ID=8, 23, 51, and 92 for the absence of the Li line in the VIMOS spectrum and the narrow FWZI 
of the H$\alpha$ line.

The V vs. V-I diagram of the \membdef\, confirmed cluster members and of the 
\nomembdef\, nonmembers   is shown in
Fig.\ref{vivtarget} (bottom panel).
 \addtocounter{table}{1} 
In Table\,\ref{membershiptab} we list members and nonmembers selected
 in PDM2007 and those identified in
this work. 

\subsection{Spectral classification\label{spectralclassification}}
Spectral types of the observed objects were determined by using the
MILES stellar library  \citep{sanc06}
whose spectra cover the range [3525-7500]\,$\AA$ at 2.50\,$\AA$ (FWHM) spectral resolution,
which is similar to that of our spectra. The library includes 985 spectra of dwarfs,
giants, and subgiants spanning a wide
range in atmospheric parameters. 
For  the present analysis we selected only spectra of  nonvariable 
stars with $-0.3<$[Fe/H]$<0.3$, i.e. with solar-like metallicity and MK spectral types 
from B1 to M6.\footnote{The 
stars HD147379B and HD111631 are classified in the Miles library as M3 V and 
 M0.5V, respectively, despite they have similar spectra. Thus we reclassified  
 the star HD147379B 
as M0.5 V.}
  
\begin{figure}
   \centering
   \includegraphics[width=9cm]{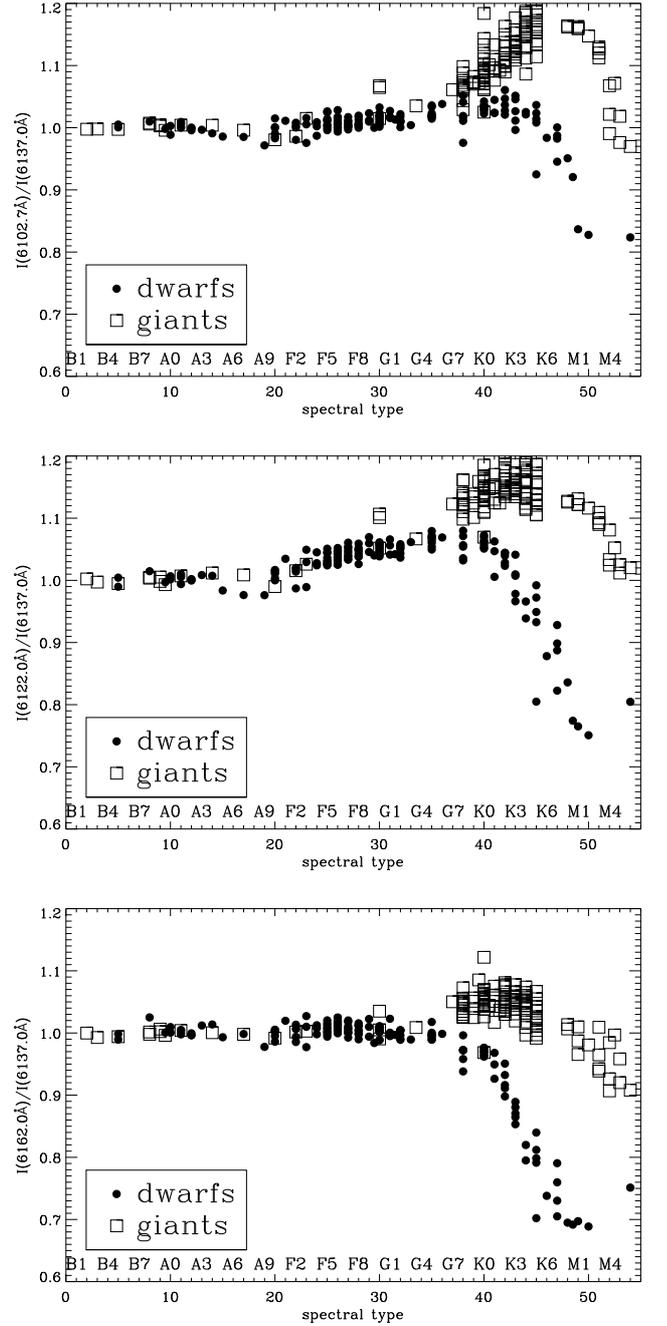}
      \caption{Ratios of the three Ca {\tiny I} line intensities at 6102.7 (top panel), 
6122.0 (middle panel), and 6162.0\,$\AA$ (bottom panel), respectively,  with respect to the 
 Fe {\tiny I} line at 6137\,$\AA$,
      used to determine the luminosity class of our targets.}
         \label{cafe}
   \end{figure}

The properties of lines in  stellar spectra are very sensitive to the luminosity class,
and thus the spectral classification, especially 
 for contracting YSOs with expected low gravity, 
 can be reliable only if we compare spectra 
with the same luminosity class.
To this aim 
we used the Ca {\tiny I} triplet lines (6102, 6122, 6162\,$\AA$) 
since these lines are very strong in giants, and  the wing profile is sensitive to
the surface gravity \citep{edva88}. In particular, we computed the 
ratios of the intensity of the three Ca {\tiny I} lines  over the 
intensity of the nearby  Fe {\tiny I} line at 6137\,$\AA$ by using spectra normalized to the continuum
 in a region of 40\,$\AA$
around each line.
 Figure\,\ref{cafe} 
shows these  ratios for the standards as a function of
spectral type  for the giants and main sequence stars. 
The plots show that these ratios can be used as diagnostics of
luminosity class as for objects later than G7.

We computed the same ratios
for the   spectra of our targets and found that only the stars with ID=74, 82, and 89
are compatible with the trend of giants, while all the other objects can be classified as
main sequence stars. 
   \begin{figure*}
   \centering
   \includegraphics[width=\textwidth]{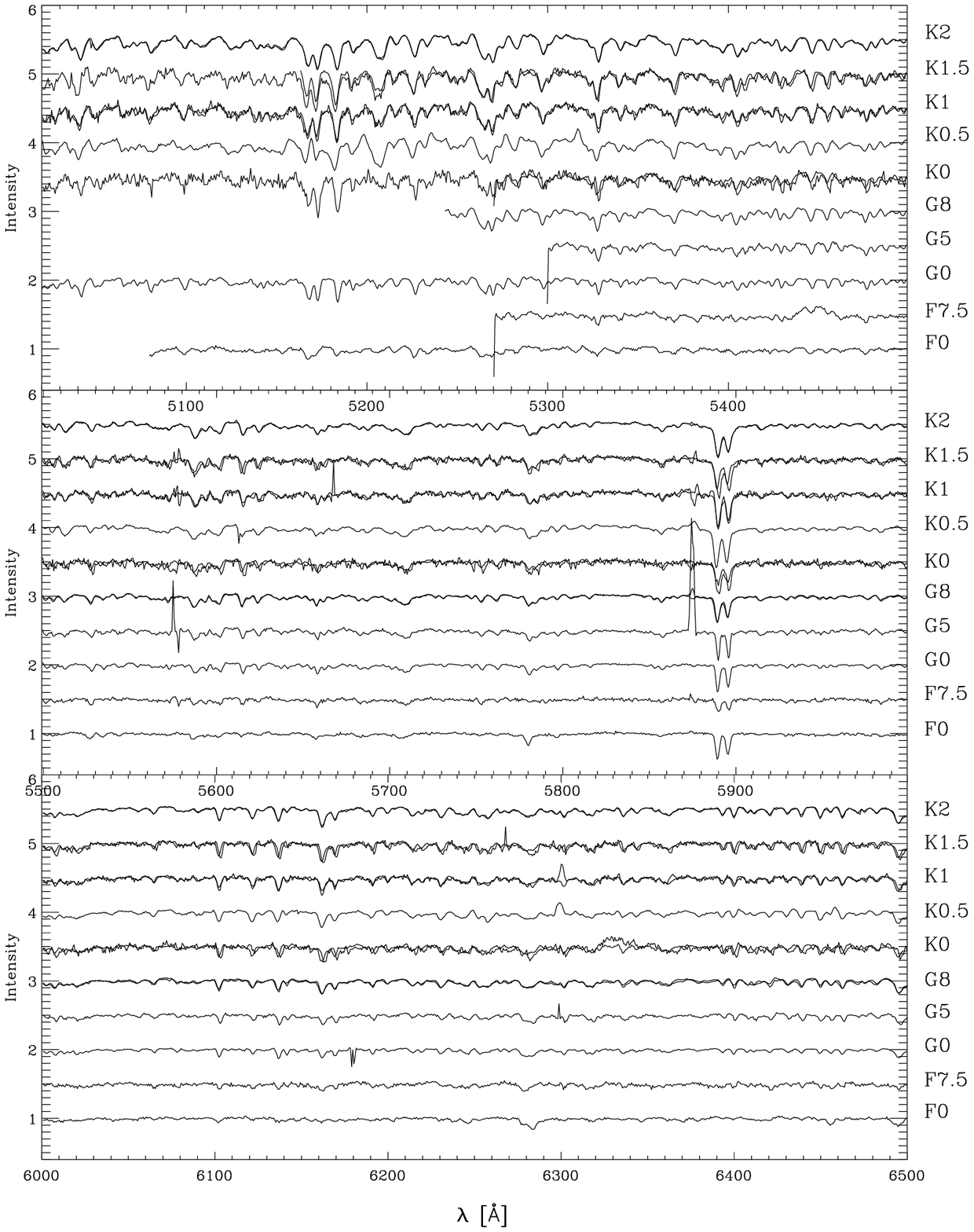}
      \caption{Spectra of cluster members with spectral type between F0 and K2,
 in the region 5020-6500\,$\AA$,   
normalized to the continuum and arbitrally shifted for clarity. }
         \label{comparespec}
   \end{figure*}

   \begin{figure*}
   \centering
    \includegraphics[width=\textwidth]{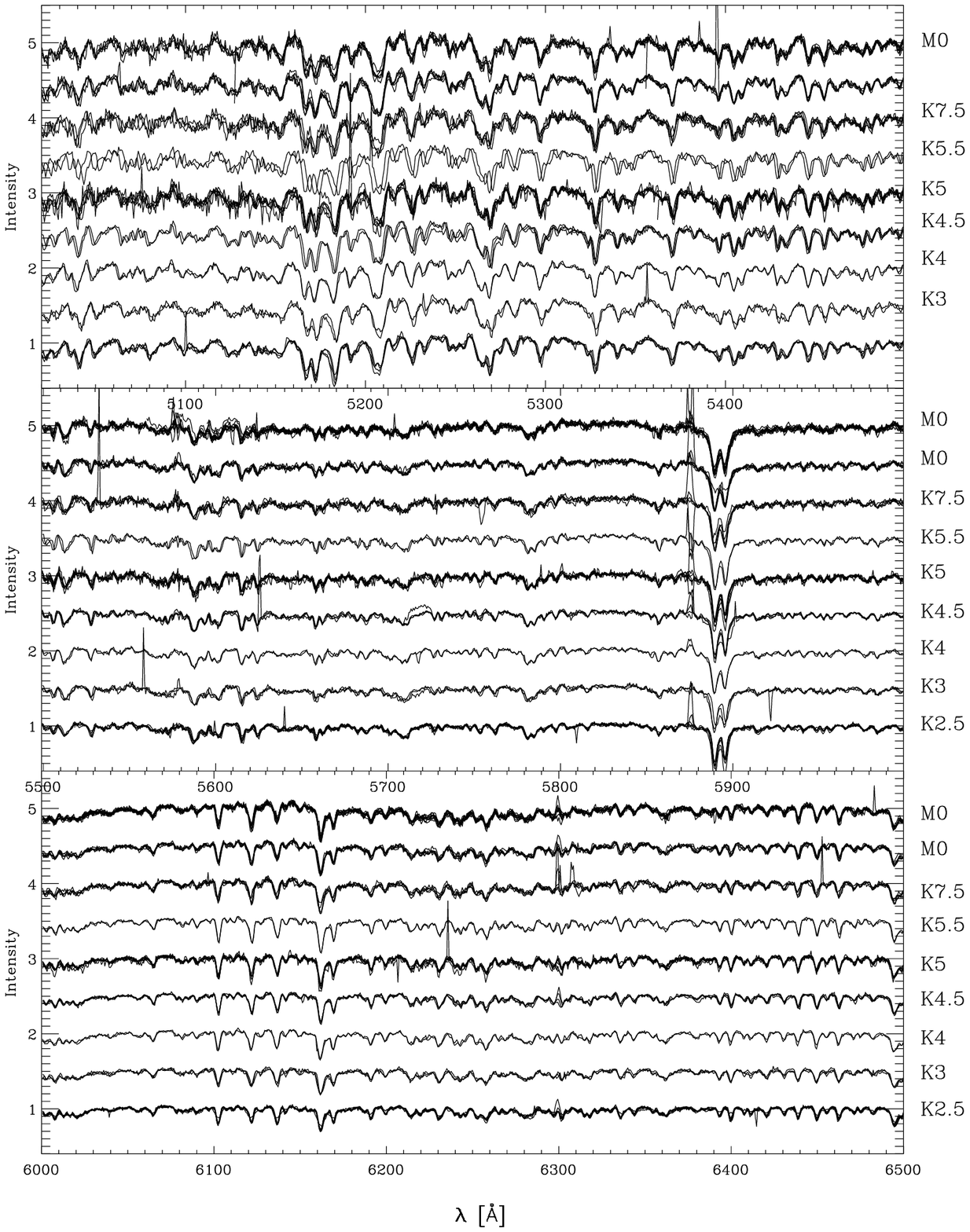}
      \caption{Spectra of cluster members with spectral type between K2.5 and M0,
 in the region 5020-6500\,$\AA$,   
normalized to the continuum and arbitrally shifted for clarity.}
         \label{comparespecK}
   \end{figure*}
   \begin{figure*}
   \centering
  \includegraphics[width=\textwidth]{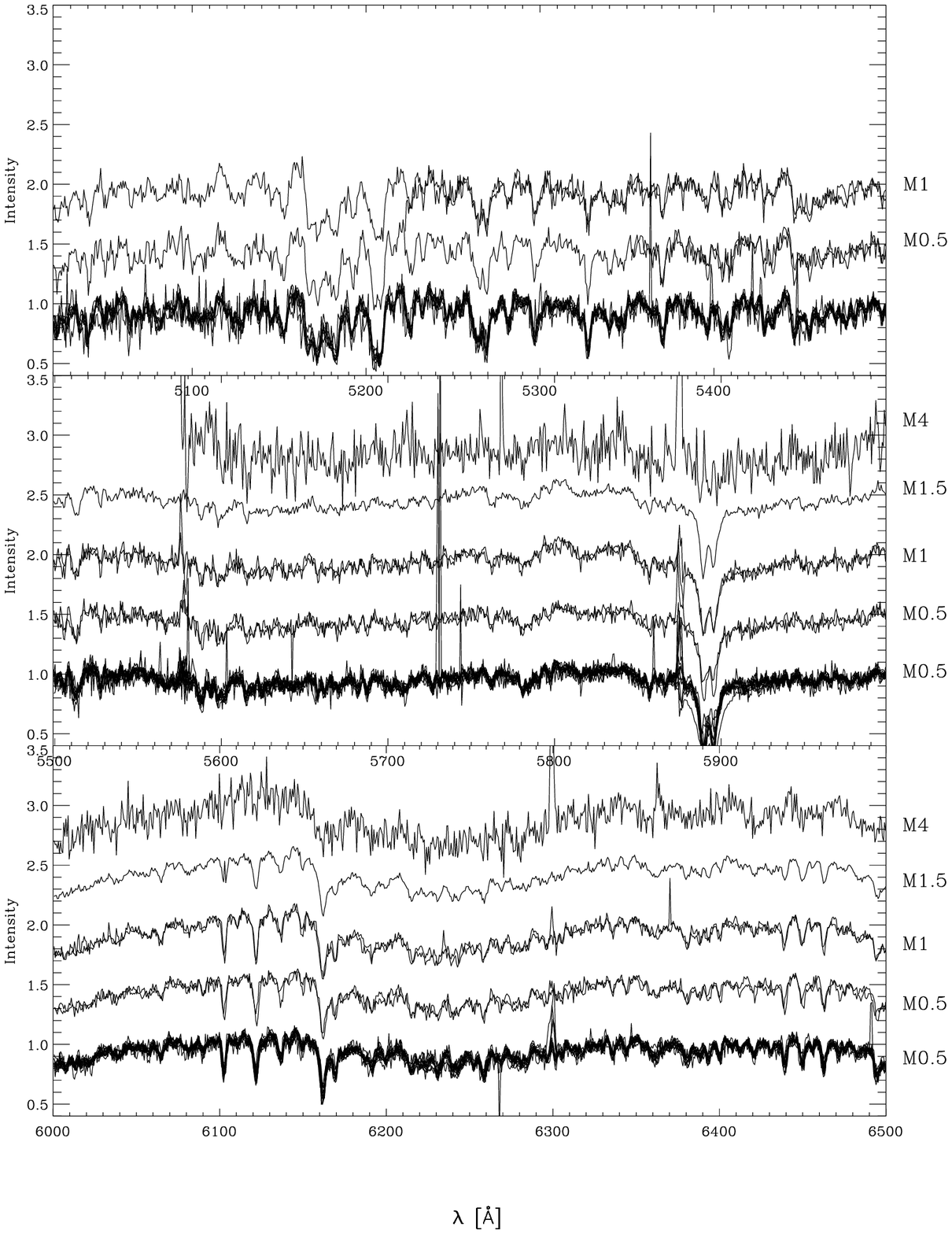}
      \caption{Spectra of cluster members with spectral type between M0.5 and M4,
 in the region 5020-6500\,$\AA$,  
normalized to the continuum and arbitrally shifted for clarity.}
         \label{comparespecM}
   \end{figure*}

   \begin{figure*}
   \centering
  \includegraphics[width=\textwidth]{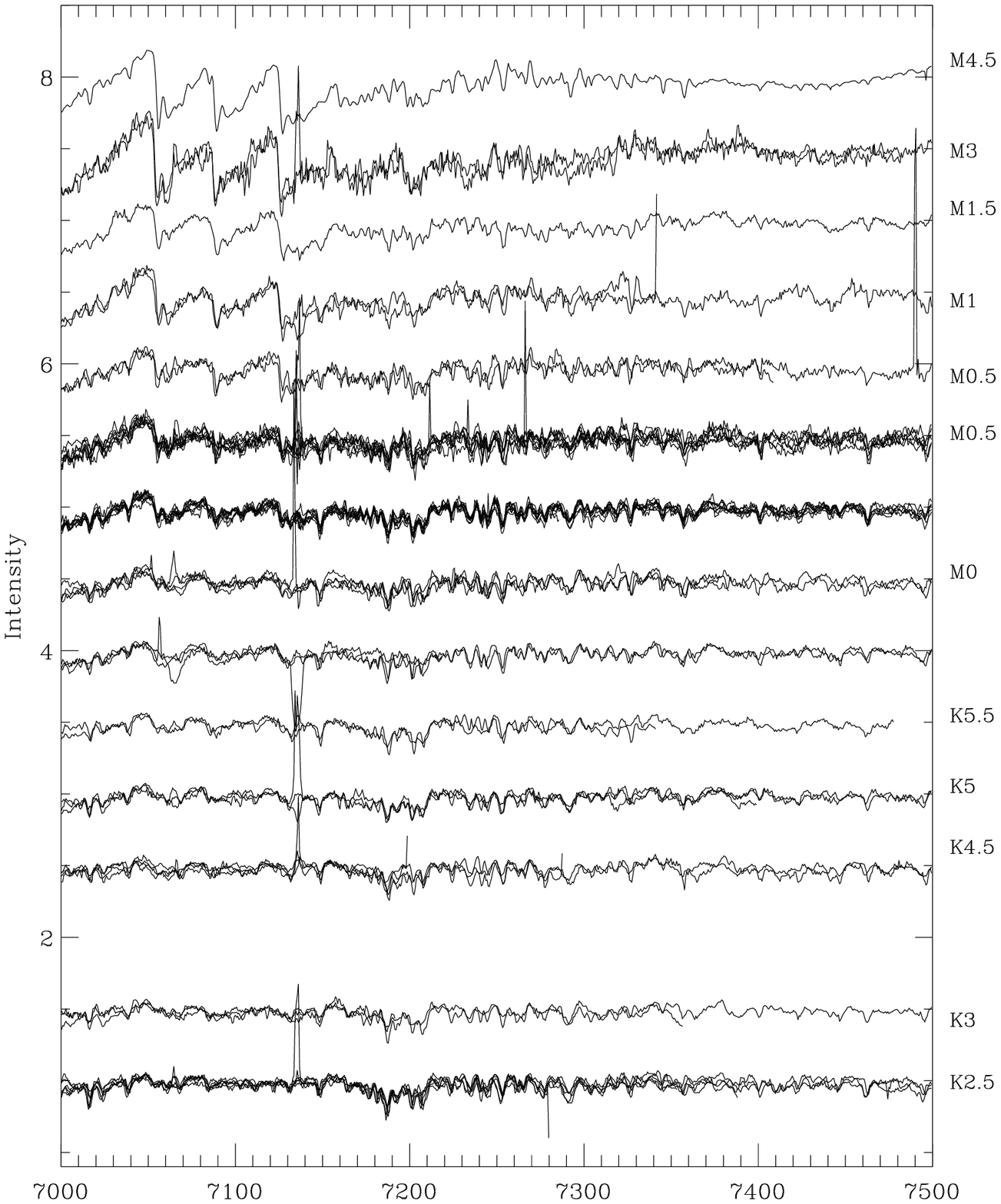}
      \caption{Spectra of cluster members with spectral type between K2.5 and M4,,
 in the region 7000-7500\,$\AA$,   
normalized to the continuum and arbitrally shifted for clarity.}
         \label{comparespecKM}
   \end{figure*}

Spectral  classification is usually based on the comparison
of   lines  forming in the stellar atmosphere, but in 
 PMS stars  the presence of additional lines originating in the circumstellar
or interstellar material implies one  difficulty in  determining spectral types.
Thus the comparison of our spectra with the standard ones has been made by excluding
the following regions (which include also several telluric lines): 
(a) 6270--6300\,$\AA$, around the O$_2$ telluric line at 6277\,$\AA$ \citep{cont95}, 
(b) 6850--6930\,$\AA$, and 7570--7700\,$\AA$, including the  O$_2$\,B and A bands \citep{jao08}, 
(c) 7150--7220\,$\AA$, including the H$_2$O absorption \citep{sanc06},
(d) 5773--5800\,$\AA$, including the diffuse interstellar features at 5780 and 5797\,$\AA$
\citep{wall87},
(e) 5875--5905\,$\AA$ around the Na {\tiny I} D$_1$ D$_2$ lines at 5890, 5896\,$\AA$, 
which may be produced also in the circumstellar environment. Finally we also excluded
the H$_\alpha$ line and 
the region 6702--6712\,$\AA$ around the Li {\tiny I}, which is very strong only in
YSOs, while it is not present in the standard spectra. We did not consider those emission
lines typical of YSOs (see column 3 in
Table\,\ref{trendlines}), 
 since not all spectra show them, and they can
originate from the hot circumstellar material and/or from the surrounding nebula.

Spectral comparison has been made
by splitting each spectrum in the spectral regions
5020--5500, 5500--6000, 6000--6500, 6600--6850, 6930--7550, and 7750--8100\,$\AA$\footnote{Note, which 
the  spectra of our targets  cover a variable  spectral range between $\sim$5000 and
 $\sim$8100\,$\AA$},
which were normalized to the continuum with IDL 
by using the same parameters (an order 3 polynomial function) 
for both  the standards and our targets. For each 
target and for each spectral range, we computed 
the difference between the target spectrum and the standard
ones to compute the residuals.
Classification has been established
by considering the spectral type of the standard giving
the minimum rms of the  residuals. Results have been checked by a visual inspection 
of the residuals.  

The normalization can be problematic for stars later than K7\,V, where
the TiO and VO molecular bands  appear to be deeper towards later spectral types.
However, since we used the same parameters and the same spectral ranges, 
 the resulting continuum normalization
of target and standard spectra can be used in order to estimate the difference
among the spectra. Uncertainties were assigned on the basis of the goodness of the comparison 
with the standard spectra.

This method is efficient to classify K and M stars whose spectra show several lines
but give very large uncertainties for earlier type stars.
Thus, for the few objects (10) 
with spectral type earlier than K, we used a second method based on the
intensities of the lines given in Table\,\ref{trendlines} as a function of the
spectral types of the standar stars.
We  measured the intensity of these lines 
on the spectra normalized to the continuum by using 20 spectral ranges 
of width from  30 to 50\,$\AA$   around the adopted lines. 
For each line and for each target a spectral type has been estimated by comparison
with the trend of the intensity obtained for the standards. The final spectral type of each 
target has been established as the most recurrent value by considering all the adopted lines.

\begin{table}[!h]
\caption{Typical lines found in our target spectra \label{trendlines} }
\begin{tabular}{cccc}
\hline\hline                 
Absorption lines  & $\AA$ & Emission lines & $\AA$ \\    
\hline                        
            	Fe\,{\tiny I}     		&    5079.0   &   O\,{\tiny I}   & 6300.3   \\       
		Fe\,{\tiny I}+Mg\,{\tiny I}   	&    5167.0   &   O\,{\tiny I}   & 6363.8 \\
               	Fe\,{\tiny I}+Mg\,{\tiny I}  	&    5172.7   &   S\,{\tiny II}  & 6716.0 \\    
               	Mg\,{\tiny I}    		&    5183.6   &   S\,{\tiny II}  & 6731.0\\     
        	Fe\,{\tiny I}+Ca\,{\tiny I} 	&    5270.0   &   N\,{\tiny II}  & 6583.0 \\     
               	Fe\,{\tiny I}     		&    5328.0   &   N\,{\tiny II}  & 6548.0 \\     
               	Fe\,{\tiny I}     		&    5406.0   &   N\,{\tiny II}  & 6527.0\\     
		Mg\,{\tiny I}			&    5528.3   &   He\,{\tiny I}  & 5875.0 \\
              	Ca\,{\tiny I}     		&    5589.0   &   He\,{\tiny I}  & 6678.0 \\     
                Fe\,{\tiny I}    		&    5615.5   &   He\,{\tiny I}  & 7065.0\\     
              	Fe\,{\tiny I}    		&    5709.5   &   Ar {\tiny III} & 7135.0\\     
               	Ca\,{\tiny I}    		&    5857.5   &   &\\     
               	Na\,{\tiny I}    		&    5890.0   &   &\\ 
		Na\,{\tiny I}    		&    5896.0   &   & \\    
               	Ca\,{\tiny I}    		&    6102.7   &   &\\     
               	Ca\,{\tiny I}    		&    6122.0   &   & \\      
               	Fe\,{\tiny I}    		&    6137.0   &   & \\      
               	Ca\,{\tiny I}    		&    6162.0   &   & \\     
\hline  
\end{tabular}
\end{table}

The results of our spectral classification are given in Table\,\ref{membershiptab}.
Figures\,\ref{comparespec}, \ref{comparespecK}, \ref{comparespecM}, and \ref{comparespecKM} show
 the spectra of cluster members
normalized to the continuum in the spectral regions mainly used for the spectral classification.
 For each spectral type, spectra have been
 arbitrally shifted for clarity. Objects of the same spectral type are shifted by the same
amount. In some cases, emission lines originating in the Nebula and/or the circumstellar disk
can be found in absorption. This is likely due to an overestimate  of the sky subtraction for these objects.

\begin{figure}
   \centering
  \includegraphics[width=9cm]{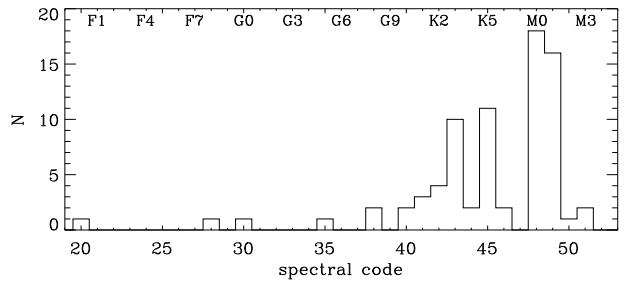}
      \caption{Histogram of the spectral types found by using the VIMOS spectra.}
         \label{histspectype}
   \end{figure}

Figure\,\ref{histspectype} shows the histogram of the spectral types derived in this work.
The sample is dominated by K and M stars, which total \nkstars\, and \nmstars, respectively.
Among the cluster members, we were not able
to classify the Class\,0/I star with ID=31, since the spectrum show only emission lines and star with ID=46,
which it is likely  a high rotator, so it shows very few lines. 

%

\subsection{Veiling}
Spectra of YSOs can be affected by veiling, i.e. a contribution to the continuum, 
likely from the  material falling towards the inner star along columns beamed by 
the magnetic field. The total accretion column emission can be described by an optically thick
emission originating in the heated photosphere and by an optically thin emission due to 
the infalling material \citep{calv98}. 
For the aim of our work, in case of veiling we can have two effects: first, a change
 in the photospheric optical magnitudes
owing to an excess in the blue part of the spectrum; and second, a wrong  spectral type determination 
since in the case of veiling, lines can appear weaker. This can lead to attributing an
earlier spectral type than the true one.

A detailed analysis of the possible role of the accretion in the BVI bands was  
conducted by \citet{dari10}, who simulated a typical accretion spectrum and added it to 
a reference model for the photospheric colors of YSOs in the Orion Nebula Cluster.
They calculated the displacement in the V-I vs. B-V diagram caused by the presence of ongoing
accretion assuming different ratios of the accretion luminosity with respect to the total luminosity.
Their results show that the effect is greater on the B-V than on the V-I and that
even an accretion luminosity lower than 10\% of the total luminosity of
the star has a strong effect on the colors of low-temperature
stars.
 
\begin{figure}
   \centering
  \includegraphics[width=9cm]{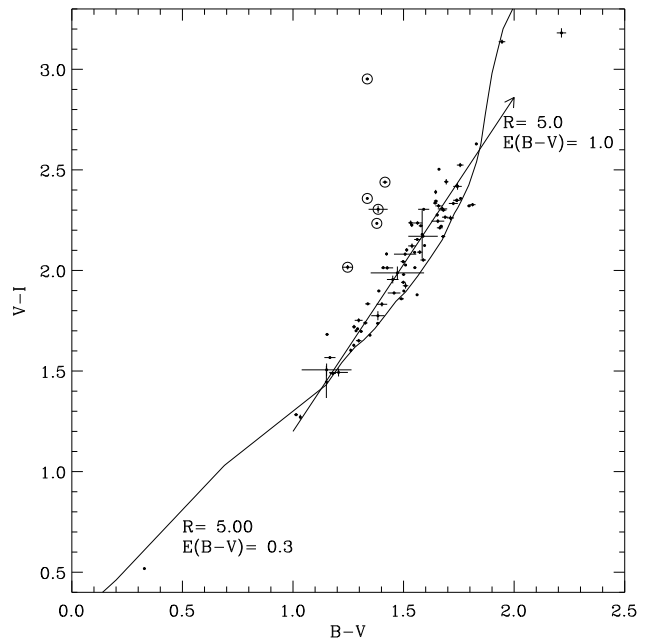}
      \caption{ V-I and B-V diagram of NGC6530 members presented in this work. Solid line
 is the locus of photospheric colors from \citet{keny95}  
reddened by using the mean cluster reddening E(B-V)=0.35 \citep{sung00} and the 
\citet{muna96} reddening law for R=5.0 (see Sect.\,\ref{reddeninglaw}) . The reddening 
vector assuming E(B-V)=1 is also drawn. Empty circles indicate objects with evident excess in the
optical colors. }
         \label{bvi}
   \end{figure}

To highlight the presence of possible objects with large veiling in our sample,
we compared the V-I and B-V colors of our targets
with the locus of photospheric colors from \citet{keny95} of
main sequence stars reddened by using the mean cluster reddening E(B-V)=0.35 \citep{sung00},
as shown in
Fig.\,\ref{bvi}, and the \citet{muna96} reddening law for R$_{\rm V}$=5.0 (see 
Sect.\,\ref{reddeninglaw}), since this is the one that follows the bulk of our data better.  

From this plot we find that
  six objects 
show an evident excess in the optical colors, as indicated in Table\,\ref{membershiptab}. Three of them
(ID=4, 74, 93) also show a broad H$\alpha$ line while 4 were classified by us as class\,II YSOs
for their NIR excesses.  For these objects we have strong indicators of  membership with
evidence of accretion, but
we cannot estimate reliable photospheric parameters, so
they will be considered with caution in the following analysis. 

Figure\,\ref{bvi} also shows that the bulk of our data follows the reddening direction 
and, within the possible spread due to the uncorrect reddening, they 
 do not show any strong excesses with respect to the expected photospheric colors. Thus we 
can consider the magnitudes of these objects representative of their photosphere
 and can be used to derive reliable stellar parameters (see Sect.\,\ref{reddeninglaw} 
for a detailed discussion). 

As already mentioned, a further effect of a possible veiling is an unrealistic spectral type
classification. Nevertheless, as described in Sect.\,\ref{spectralclassification}
for each objects, we  split the whole spectrum into more regions and 
verified that the adopted  final
spectral type is consistent in every subspectrum, by considering that some spectral regions
are more  sensitive to veiling than others, depending on the spectral type. 
In particular, since most of our targets are late K and M stars, we mainly used the  reddest
part of the spectrum -more sensitive for these spectral types-,  which is expected to be
less affected by veiling \citep{gull00}. In addition, since most of our targets are
candidate class\,III YSOs, with no strong evidence of a circumstellar disk, we can conclude
that our spectral classification should not be influenced by the veiling, with the exception
of the star with ID=31, which we were not able to classify since its spectrum 
 shows only emission lines. This object was classified as class\,0 by \citet{kuma10}
by using IRAC--Spitzer photometry.

In addition we find that star  ID=86 shows  a very peculiar spectrum with 
 the reddest region
(6900-7400\,$\AA$)   similar to that of K7-M0.5\,V star,   the region 6000--6500\,$\AA$ 
  similar to a  K3-K4\,V star, while   the region 5000--5500\,$\AA$   similar to  
a G8\,V star. Since the FWZI
is quite high (~19\,$\AA$), we guess that this object is affect by veiling and/or variability.
Given its previous spectral classification (B8.5\,V from SIMBAD database) and its bright magnitude,
we do not exclude this object being a candidate FU-Ori object or
a binary where we can see the low-mass companion spectrum.
Based on inconsistent features in their spectra, 
we also suspect a possible effect of veiling for the stars   ID=12 and 36.  
Even in this case, we consider these objects with caution for the following analysis.

\subsection{Intrinsic colors and reddening \label{colorsredd}}
   \begin{figure}
   \centering
  \includegraphics[width=9cm]{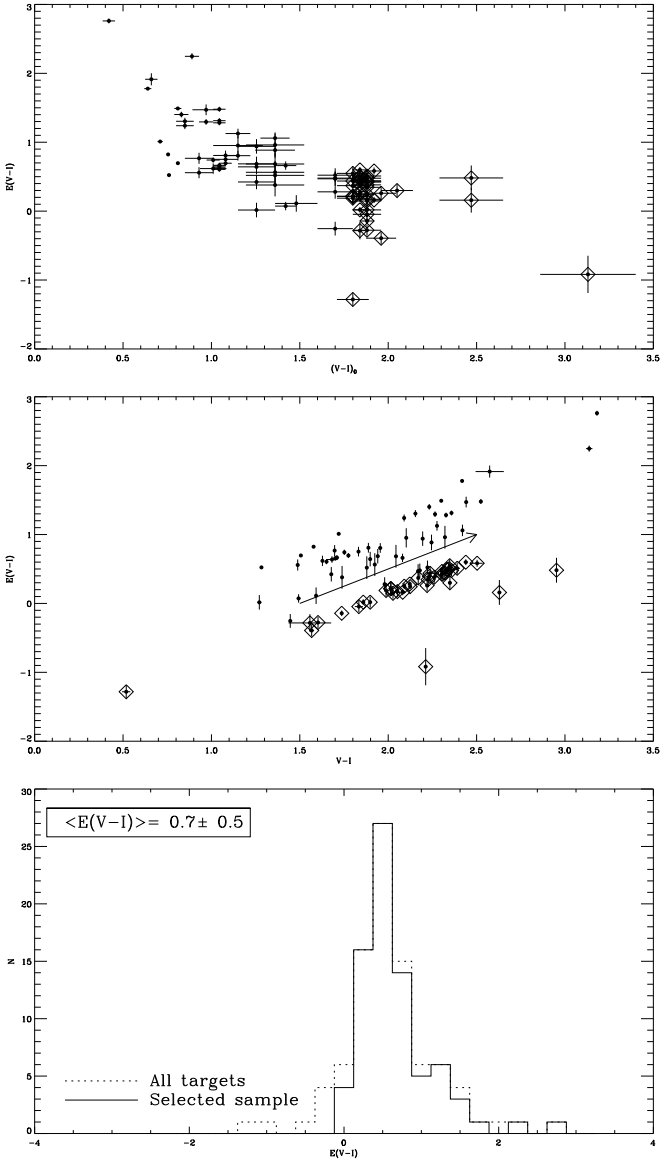}
      \caption{E(V-I) vs. intrinsic (V-I)$_0$  (top panel) and vs.  observed (V-I)  (middle panel).
 Objects indicated by empty diamonds are M0 or later stars. The reddening vector corresponding to 
E(V-I)=1 is also plotted. E(V-I) distribution (bottom panel) obtained by considering all targets (dotted line)
and the selected sample including the 70 cluster members with E(V-I)$>0$ and without special photometric
or spectroscopic features indicated in Table\,\ref{membershiptab}.}
         \label{intrinsiccolors}
   \end{figure}

To convert spectral types into intrinsic colors,  
 we used the \citet{keny95} transformations. 
  We verified that for the colors of our targets, the \citet{keny95}  relation
 is very similar to that suggested in \citet{luhm99}, which considers the \citet{keny95} 
 colors for  types earlier than M0 and the colors from \citet{legg92} for M0 and later types.
Figure\,\ref{intrinsiccolors} shows individual reddenings E(V-I), derived from the difference between
the observed and the intrinsic colors, as a function of the intrinsic (V-I)$_0$  (top panel)
and of the observed (V-I) (middle panel). 

In Fig.\,\ref{intrinsiccolors}   we also have unphysical negative reddening  values
for a few members. We find that the seven YSOs with E(V-I)$<$0 are 
all YSOs of spectral type
equal or later than M0, and two of them show
a quite high FWZI of  H$\alpha$ ($>15\AA$). We guess that these negative
values are due to observed colors that are bluer than the pure photospheric ones for a possible
 excess in the V band owing to accretion processes.
This is expected for late type stars, since, as shown in \citet{dari10},
low fractions of accretion
luminosity have a strong effect on the colors of low-temperature
stars.

\begin{figure*}
   \centering
   \includegraphics[width=\textwidth]{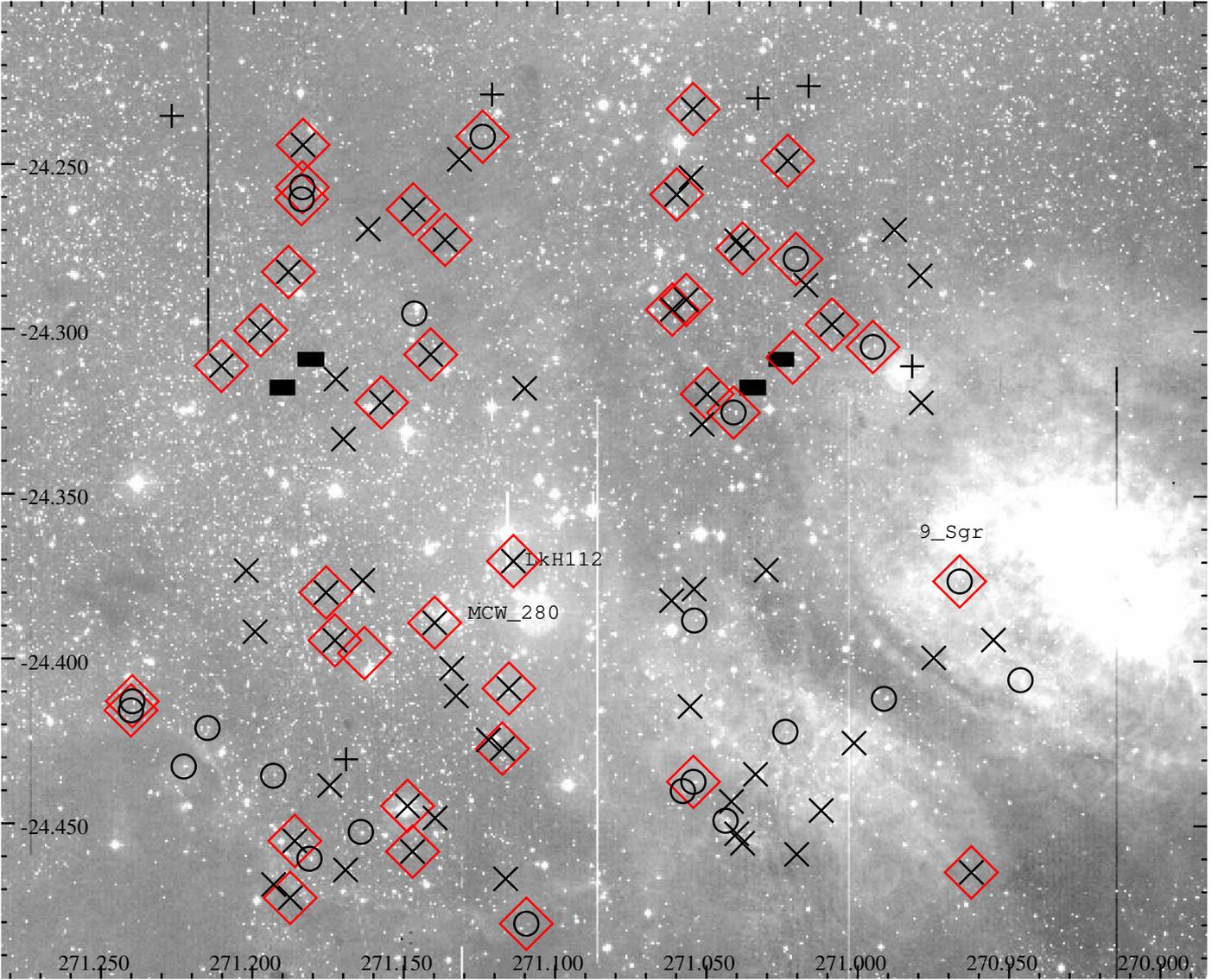}
     \caption{Spatial distribution of all the objects studied in this work overplotted
on the combination of the two deep-dithered I band WFI images \citep{pris05}.
 {\tt X} symbols indicate WTTS, circles indicate CTTS, plus indicate nonmembers, and  
 diamonds 
indicate M0 or later stars.
 O stars falling in the  field of view (FOV) 
investigated in this work
are also indicated. Coordinates RA(J2000) and Dec(J2000)
 are given  in hours and degrees, respectively.}
         \label{spatialdistr}
   \end{figure*}

In Fig.\,\ref{intrinsiccolors} (middle panel), all objects follow the reddening vector, but
the mean reddening of M stars is lower 
than the one of earlier type stars, and the two populations are quite distinct. This is a selection 
effect, since  M stars more reddened than the observed ones  would be fainter than our observational
limit.
This is  consistent with the spatial distribution of the objects studied in this work,
shown in Fig.\,\ref{spatialdistr}, where M type stars  are homogeneously distributed 
 in the N and E regions but are   
almost completely absent in the  SW region, where the mean reddening is higher and/or
the sensitivity  is lower for the strong nebular $H_\alpha$ emission.

Finally, Fig.\,\ref{intrinsiccolors} (bottom panel) shows the reddening distribution by considering the whole
sample and the subsample of 78 cluster members with E(V-I)$>0$, which  
we will indicate hereafter as the selected sample. If we consider this subsample,
we find   a mean value of E(V-I)=0.7 with an rms equal to 0.5. By using the adopted reddening laws, the mean
value corresponds to E(B-V)=0.42, which, within the errors, is consistent with the value 
$<$E(B-V)$>$=0.35 found by \citet{sung00}.

\subsection{Reddening law \label{reddeninglaw}}

To derive stellar luminosities, we also need to know the interstellar absorption,
which can be derived from the reddening values. 
This step requires assuming a suitable reddening law (R$_{\rm V}$), which, in  many star-forming clouds,
has been proved   to be different from the standard interstellar
extinction law   \citep{povi11}. In addition, very recently, \citet{fern12} has suggested 
 for NGC\,6530 an anomalous extinction law (R$_V$=4.5) by using the distribution of known E(V-K) vs. 
E(B-V) values. Their sample is, however, limited to objects with E(B-V)$\lesssim$0.4.

To constrain the reddening law adequate for this cluster, we used the ratios E(V-I)/E(B-V) and
E(V-K)/E(B-V). In fact, for weak T-Tauri stars, the observed colors should only be affected 
by interstellar absorption and the reddening ratios derived by using different colors should
be    sensitive to  the total-to-selective extinction
R$_{\rm V}$, which depends on composition and size of the grain in the interstellar medium.
For example, \citet{fitz09} suggest
 to use the relation  R$_{\rm V}$=1.36$ \frac{\rm E(K-V)}{\rm E(B-V)}$-0.79
to derive the reddening law. 

\begin{figure}
   \centering
  \includegraphics[width=9.0cm]{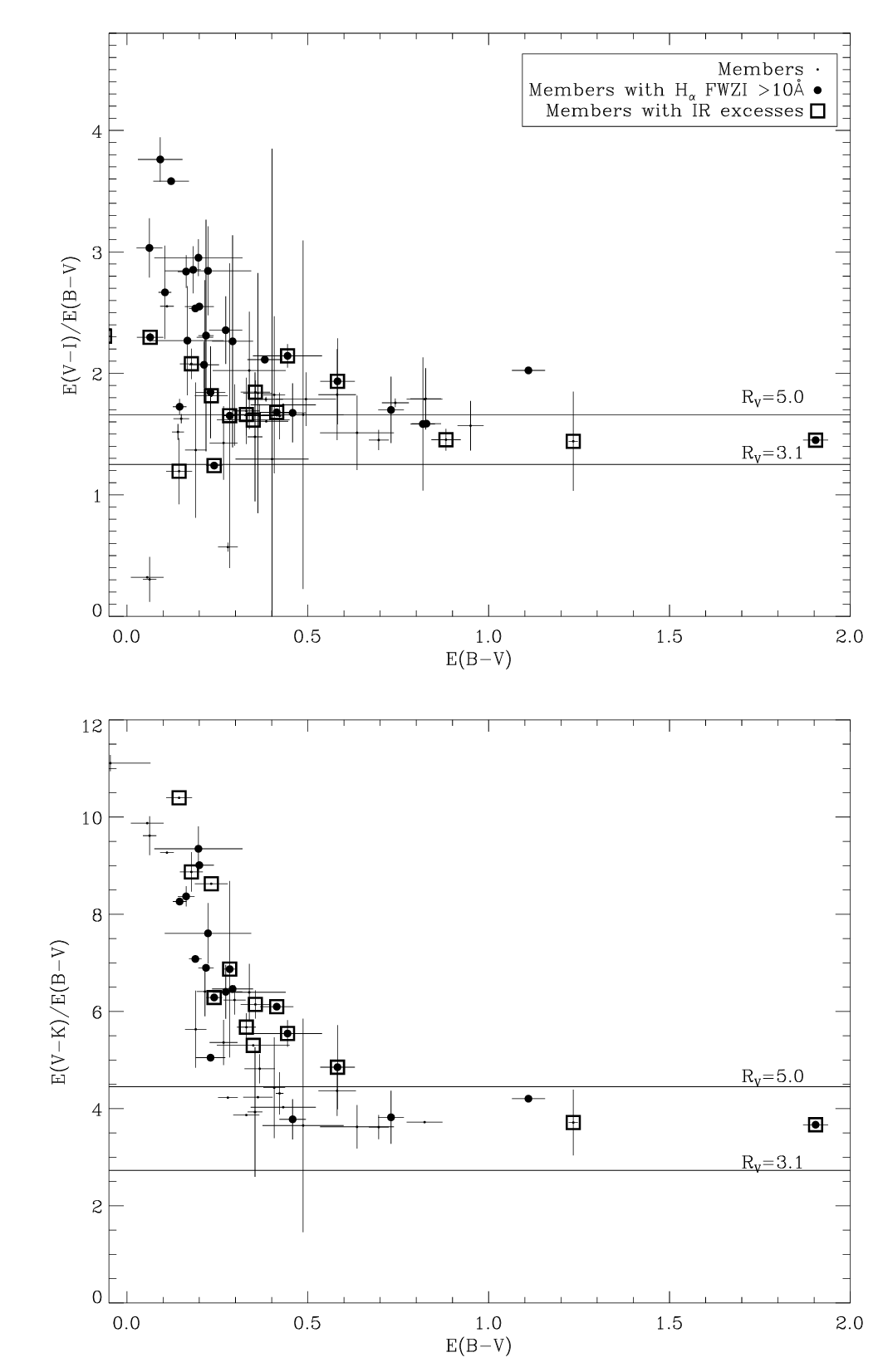}
      \caption{Color excess ratios as a function of E(B-V) computed for our targets.
Horizontal solid lines indicate the color excess ratios
	predicted by the  \citet{muna96} (top panel) and \citet{fior03} (bottom panel) reddening laws
for  R$_{\rm V}$=[3.1, 5.0].}
         \label{deriverv}
   \end{figure}

To this aim, we selected cluster members with E(B-V)$>$0. In addition, we discarded peculiar
objects with features reported in the notes of Table\,\ref{membershiptab}.
 Figure\,\ref{deriverv} shows   E(V-I)/E(B-V)  (top panel) and   E(K-V)/E(B-V) (bottom panel)
as a function of E(B-V). Candidate accretors, defined as objects with an FWZI in the
the H$_\alpha$ line higher than 10\,$\AA$, and objects with IR excesses are indicated by
different symbols.
The ratios  E(V-I)/E(B-V)=[1.25,1.66]
and    E(V-K)/E(B-V)=[2.73,4.45] 
  for R$_{\rm V}$=[3.1, 5.0], by using the \citet{muna96} 
  reddening law for the optical bands and the \citet{fior03} for the 2MASS bands are also indicated in the
  figure.
The extinction coefficients by \citet{muna96} and \citet{fior03}
 were derived by using the \citet{math90} and \citet{fitz99}
 reddening laws, respectively, weighted over the V\,R$_{\rm C}$\,I$_{\rm C}$ and J\,H\,K$_{\rm S}$ band profiles 
of the \citet{bess76} and 2MASS photometric systems, respectively. 

We find that only a fraction of  our targets
show   color-excess ratios consistent with uniform R$_{\rm V}$,
 as expected for normal reddened stars, while for 
many objects the color excess ratios show a wide spread that it is likely
due to excesses in the blue B and/or in the red K bands. 
In fact, most of the accretors have high E(V-I)/E(B-V),  while almost all the objects 
with IR excesses have high E(V-K)/E(B-V).

\begin{figure}
   \centering
  \includegraphics[width=9cm]{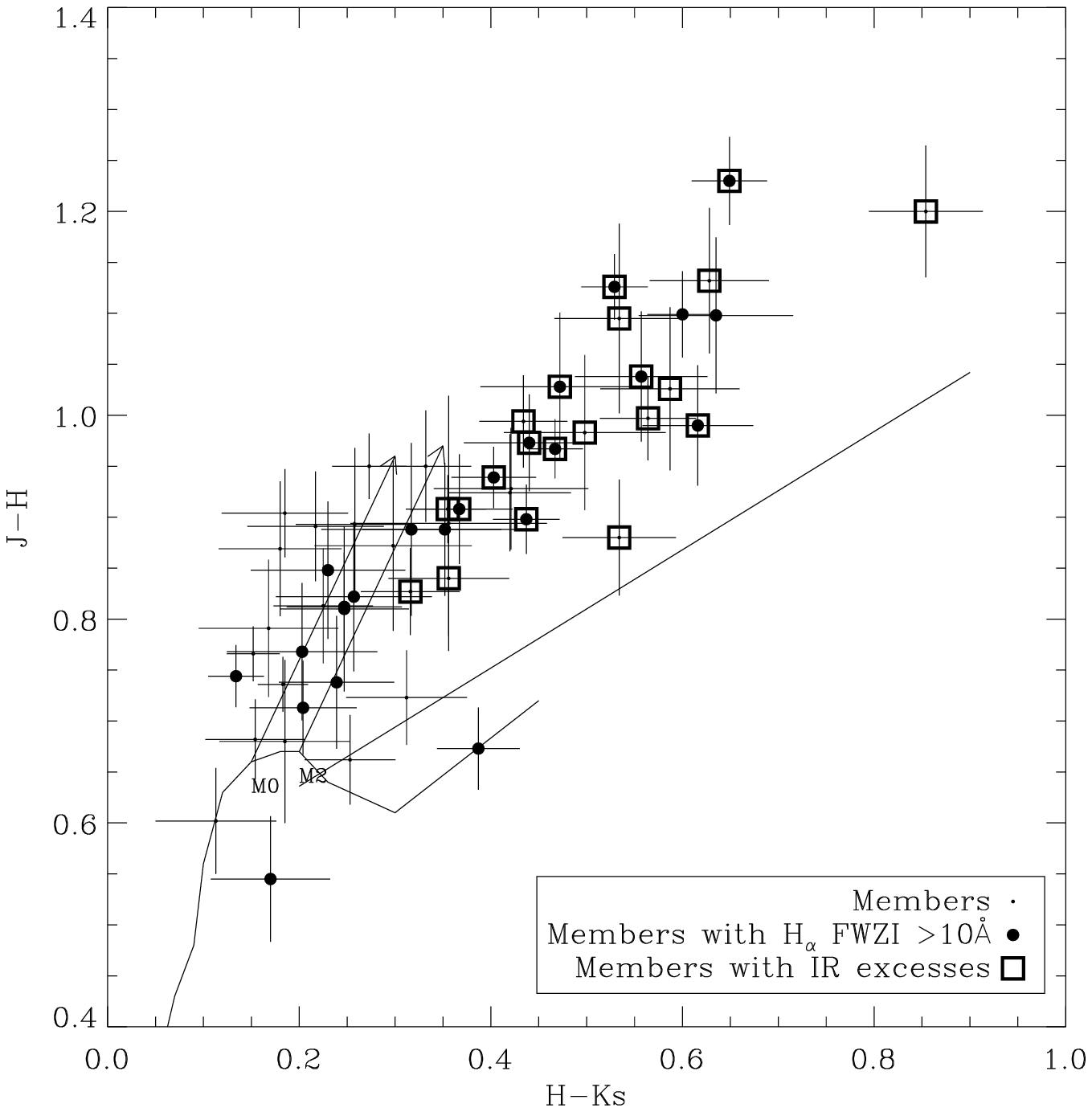}
      \caption{J-H vs H-K diagram of our targets. 
The solid curve is the \citet{keny95} locus for the colors of main sequence stars, the two arrows
indicate the reddening vectors corresponding to A$_{\rm V}$=3 starting from colors of M0 and M2 dwarfs,
and the solid line is the CTTS locus by \citet{meye97}.}
         \label{jhhk}
   \end{figure}

Objects with evident NIR excesses can also be seen in Fig.\,\ref{jhhk} where we show the J-H vs. H-K
diagram. Most of our targets show colors that are more reddened than those expected for dereddened
main sequence stars. Nevertheless, since our targets are YSOs 
of earlier spectral type than M2, the observed colors cannot be explained by   interstellar 
reddening alone,
as can be seen by the comparison with the reddening vectors.  Instead, for many of our targets,
the observed colors are consistent with the \citet{meye97} locus of CTTS stars. Therefore,
the observed colors  result  from a combination of interstellar reddening and NIR excesses from dust
from the circumstellar disk.

However, if we consider the objects with constant color excess ratios we find that the values E(V-I)/E(B-V)
are consistent with the ratio given by \citet{muna96} for R$_{\rm V}$=5.0, while if we consider the objects
with E(K-V)/E(B-V) $<4$, we find R$_{\rm V}$ in the range 4.0--4.5 by using the \citet{fitz09} equation.
Since our sample is dominated by YSOs with circumstellar disk with or without excesses, 
we cannot constrain the R$_{\rm V}$ value with high precision. Nevertheless, by considering
the consistency of the results obtained both from optical and IR magnitudes, we can conclude that
  NGC\,6530 cluster members show an anomalous extinction with 
R$_{\rm V}\lesssim $5.0. Thus for the following analysis, we use the reddening ratios given by \citet{muna96}
 for R$_{\rm V}$=5.0.

%
%
%
%
%

\section{Discussion} 
\subsection{H-R diagram, ages, and age spread \label{hrsect}}

\begin{figure*}
   \centering
  \includegraphics[width=7cm]{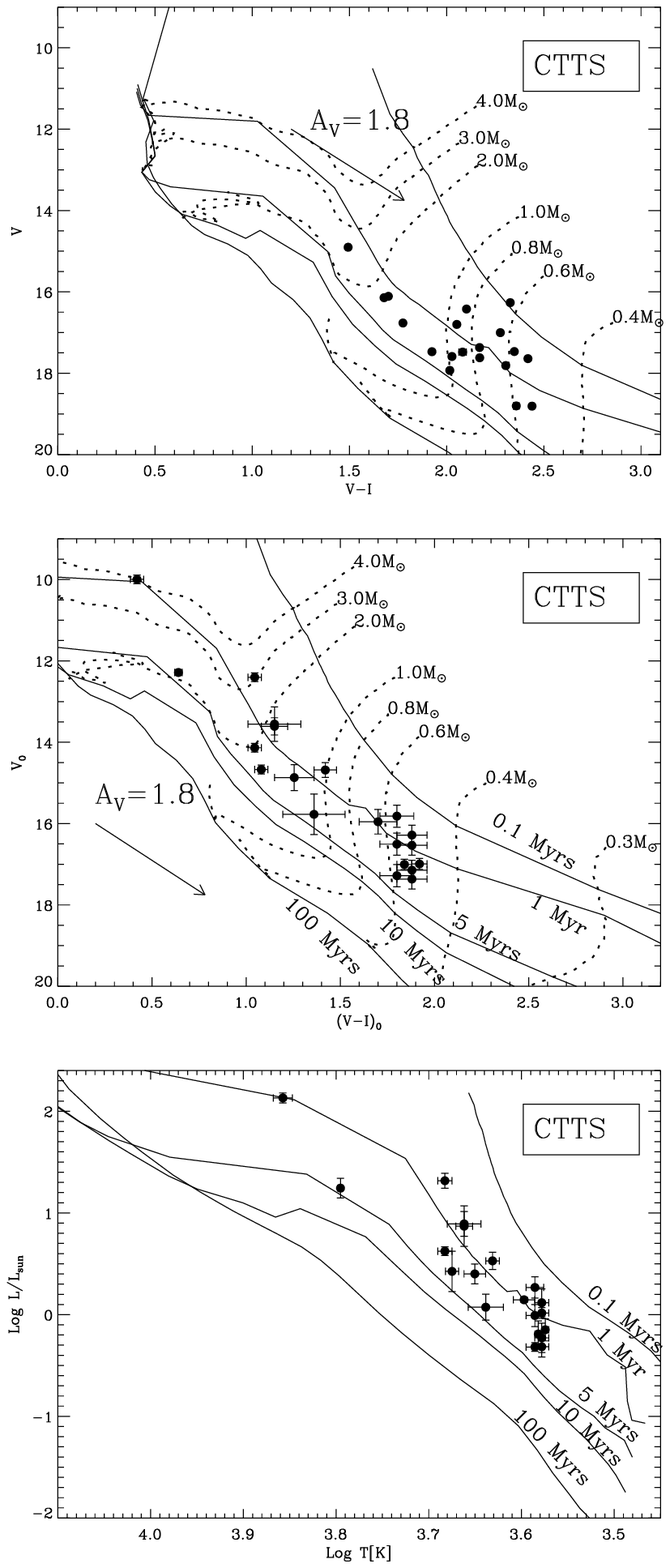}
  \includegraphics[width=7cm]{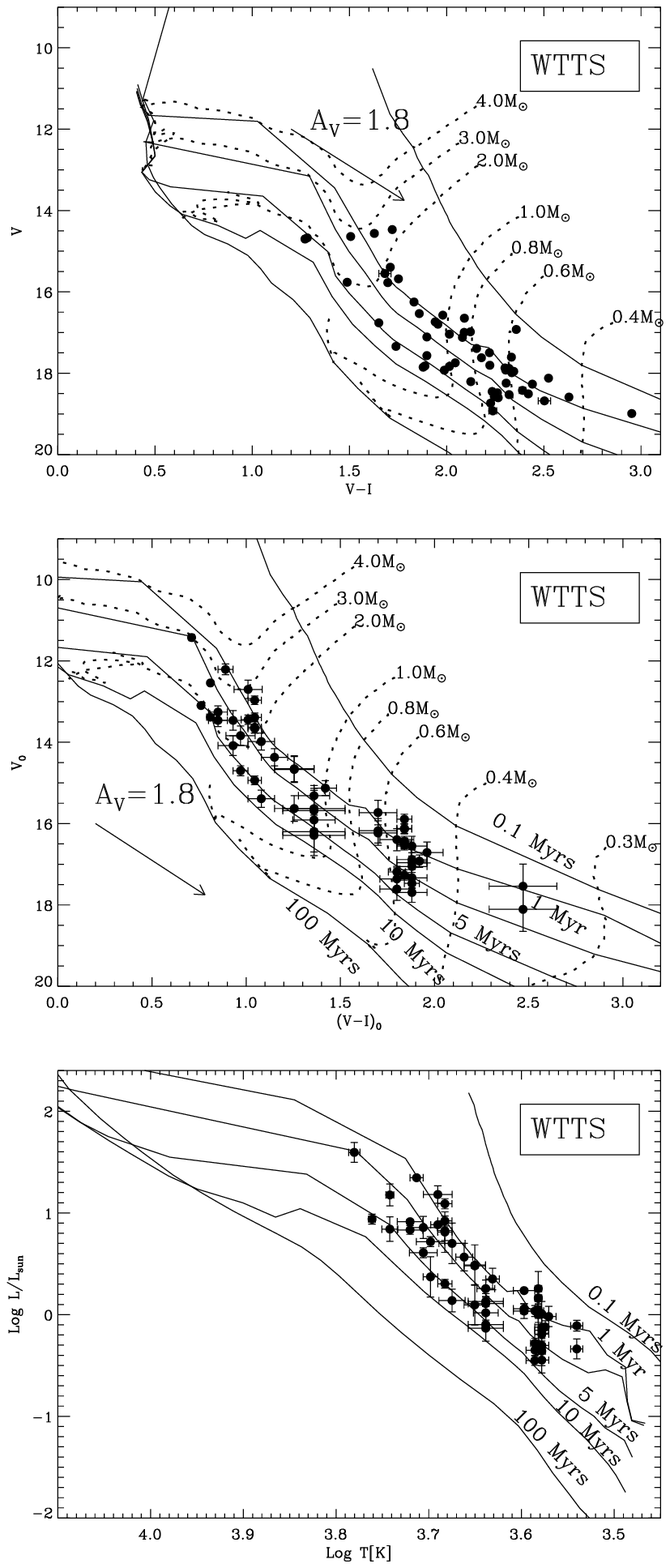}
      \caption{V vs. V-I diagram obtained by using observed (top panels) and dereddened (middle panels)
magnitudes and colors for the  CTT and WTT  cluster members with E(V-I)$>$0. 
Solid and dotted lines are the \citet{sies00} isochrones and track, respectively,
assuming the cluster distance of 1250\,pc \citep{pris05}. In the top panel, the isochrones are reddened
by assuming the literature mean reddening E(B-V)=0.35 \citep{sung00} and the \citet{muna96} reddening 
laws for  R$_{\rm V}$=5.0. In the middle panel, isochrones are only scaled for the cluster distance. The
 arrow indicates the reddening vector  corresponding to E(B-V)=0.35  for  R$_{\rm V}$=5.0. The HR diagrams
of the two samples are shown in the bottom panels.}
         \label{v0vi0}
   \end{figure*}

Figure\,\ref{v0vi0} shows the observed (top panels) and dereddened (middle panels) 
color-magnitude diagrams for the  21 CTTS (left panels) and the   57 WTTS (right panels) 
members with E(V-I)$>0$. The sample of CTTS
includes an object classified as Class\,0/I based on the IRAC--Spitzer photometry \citep{kuma10}.
The HR diagrams of the two samples are shown in the bottom panels. Effective temperatures and bolometric corrections 
were derived    by using the \citet{keny95} transformations. Bolometric luminosities 
 were determined from the V magnitudes, by considering individual absorption corrections derived by the intrinsic
 colors and  the adopted reddening law.
Masses and ages were derived by interpolating the theoretical tracks and isochrones by \citet{sies00}
to the position of the stars in the V$_0$ vs. (V-I)$_0$ diagram. 
Uncertainties 
on the stellar parameters were derived by propagating 
individual uncertainties on the spectral types.
The values of individual
extinctions,   effective temperatures,
bolometric luminosities, ages and masses, and the relative uncertainty ranges are given in 
\addtocounter{table}{1} 
Table\,\ref{finaltable}.

The comparison between the two CMDs (observed and dereddened) drawn in Fig.\,\ref{v0vi0}
shows that the actual dynamic range of the 
intrinsic magnitudes (V$_0$=10.--18.1)
is greater than that of the observed magnitudes (V=14.5--19.0). In addition, we note that the
 individual correction for interstellar
reddening allows us to reduce the apparent age spread, since we find that most of
those YSOs  that  in the V vs. V-I diagram have an apparent age in the range 0.1--1\,Myrs
are actually very reddened stars with age $\gtrsim$1\,Myrs. Finally, we see that
 the mass distributions in the observed and dereddened diagrams are very different.

\begin{figure}
   \centering
  \includegraphics[width=9cm]{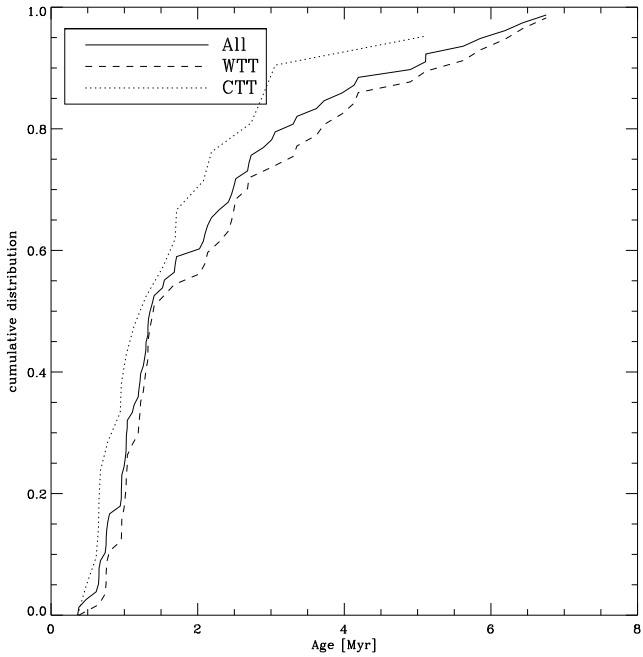}
      \caption{Cumulative age distribution of  WTTS,  CTTS and all cluster members.}
         \label{cumagectt}
   \end{figure}

Figure\,\ref{cumagectt} shows the (cumulative) age distributions
 for the WTTS and CTTS and for the
joint sample (WTTS plus CTTS). The ages of WTTS are not  significantly different from the
CTTS ones, with a marginal indication that CTTS are younger than WTTS.
As a result, we decided to consider the WTTS-plus-CTTS joint sample and found that around 60\% 
of cluster members have ages in the range $\sim$1--2\,Myrs, but there is  an
 older population with ages in the range 3--7\,Myrs. 
Even if a small part of this spread ($\sim$1.5\,Myrs) could be attributed to binaries,
we exclude the younger population
actually being a population of binaries, since in this case the two populations should
have the same spatial distribution.   

\begin{figure}
   \centering
  \includegraphics[width=8cm]{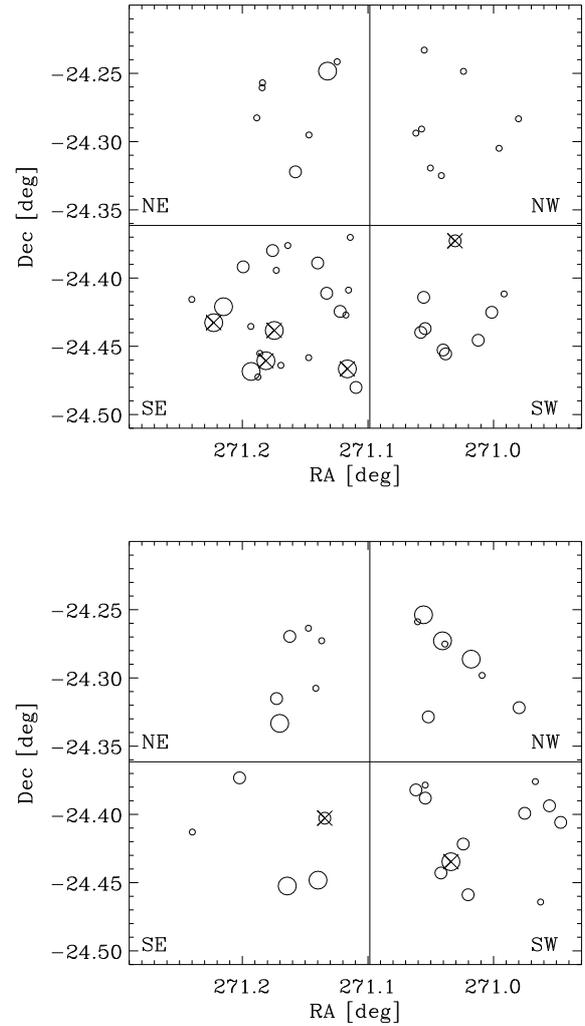}
      \caption{ Spatial distribution of the populations youngest (top panel) and oldest (bottom panel) 
than 2.0\,Myrs (empty circles). Different circle sizes correspond to objects with E(V-I)$<$0.5 (smallest circles),
0.5$\le$E(V-I)$<$1.0 (medium circles) and E(V-I)$\ge$1 (largest circles).  
 X symbols indicate objects with mass larger than 
2.5\,M$_\odot$.
The four regions  corresponding to the four VIMOS CCD quadrants are also indicated.}
         \label{agespatial}
   \end{figure}

\begin{figure}
   \centering
  \includegraphics[width=9cm]{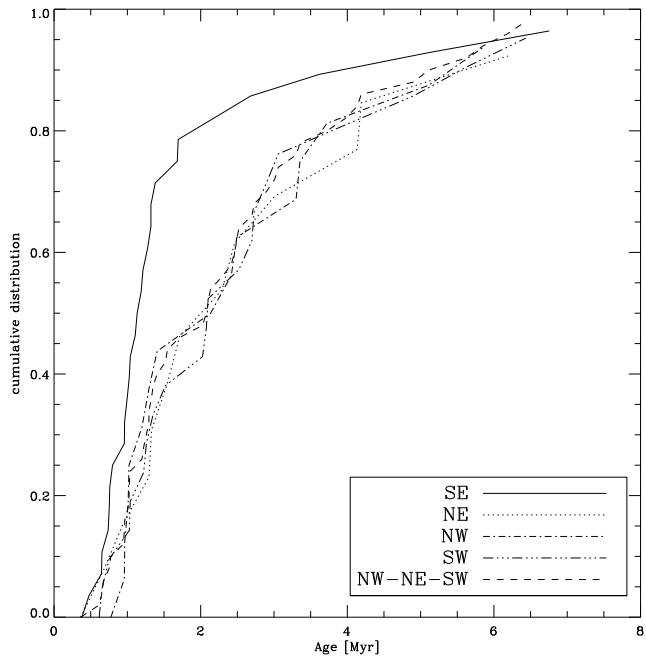}
      \caption{Cumulative age distribution of all objects in the four spatial regions shown in 
Fig.\,\ref{agespatial}. The cumulative age distributions of all the objects located in the SE and
in the other regions and of the combined SW-NW-NE sample are also shown.}
         \label{cumagespatial}
   \end{figure}

This is not the case, as shown in  Fig.\,\ref{agespatial} where we plot
 the spatial distributions of the YSOs with ages younger 
and older than
2\,Myrs falling in the N, S, W, and E
 regions corresponding to the four VIMOS CCD quadrants.   Different symbol sizes correspond to YSOs with
E(V-I) in the ranges indicated in the figure.  
 We find that, while the oldest objects are quite homogeneously located in the observed region,
about 50\,\% of the youngest objects (ages $<$2\,Myrs) are located in the  SE region.
To see whether the spatial distribution of 
the youngest population  is significantly different from the oldest one,
we derived the cumulative age distribution in the four regions shown in Fig.\,\ref{cumagespatial}.
Since the number of objects in the N  and SW region is very small, we also consider
the sample including all the objects in these three regions with the aim to have more robust statistical
results. From these distributions, it is clear that  in the SE region
  around 80\% of the objects are younger than 2\,Myr, while this percentage is much lower ($\sim$40--50\%)
in the other regions. Using the two sample Kolmogorov-Smirnov tests, we find that
the population of objects
in the SE is different from the one in the SW-NE-NW regions at a significance level of 99\%.
Figure\,\ref{agespatial} also shows that there is no correlation between the spatial distribution 
of the individual reddenings and the one of the derived ages, so we can discard the observed 
"age spread" as due to an artifact from  the treatment of the extinction.

Our results support the  scenario of sequential star formation since we find a first
generation of stars in the N regions and in the SW region
and a second generation of stars that is quite concentrated in the  SE region. Since 
our observations do not cover the central region, we cannot say anything about this
interesting region. 

Since the E(V-I) median value $\pm$ 1 $\sigma$ is equal to 0.52$_{-0.28}^{+0.52}$,
corresponding to A$_{\rm V}$= 1.57$_{-0.83}^{+1.57}$, our observational limit implies a mass limit
equal to 0.6$_{-0.1}^{+0.6}$\,M$_\odot$ at 4\,Myrs,
 not including outliers that can have very different values of
E(V-I) and  mass limit.
For these reasons, we are not able to  discuss
the sequential process as a function  of stellar masses, even if we find
that almost all seven observed YSOs with masses higher than 2.5\,M$_\odot$  are  located 
in the SE region.

Our finding  allows us to suppose that (within the FOV studied in this work)
the SE region is the most recent site of star
formation of low and intermediate mass stars where the population
 has not yet undergone (or just started)  mass segregation. In contrast,
the oldest population
 is more loosely clustered in the surrounding N and  SE
regions. 

\section{Summary and conclusions}
We used the VIMOS@VLT spectra of 94 candidate cluster members of the young open cluster NGC6530
with the aim to obtain the spectral classification, crucial for deriving accurate stellar 
parameters. First of all, we considered available membership  indicators 
(X-ray detection, radial velocities,
lithium line, and H$_\alpha$ shape) both from literature data
and/or  from our new spectra to find members. 
Our spectral resolution does not allow us to measure the equivalent width of the lithium line,
but since this line is expected to be  very  prominent in YSOs, 
we were able to use it effectively  to distinguish cluster members. 

Owing to the strong and spatially variable 
H$_\alpha$ emission from the surrounding nebula, we cannot quantify the
H$_\alpha$ emission of the YSOs that are undergoing accretion, so we were not able
to estimate the mass accretion rates. Nevertheless, we can  distinguish strong accretors
by estimating the FWZI of the H$_\alpha$ line that for these objects is larger than 
H$_\alpha$ emission coming from the nebula. Based on the available membership indicators,
 we find that 88 of our targets are
confirmed cluster members, and the remaining six are nonmembers.

Based on the comparison with standard spectra in the literature, we classified our targets and found that 
the cluster member  sample (88 YSOs) is dominated by late type 
stars with 42 K type and 32 M type stars, while for two of them
we were not able to assign a spectral type.

By considering the \citet{keny95} transformations, we derived intrinsic colors (V-I)$_0$, hence the interstellar 
reddening E(V-I). We used the V-I colors, since they should be less affected by blue excesses due to accretion
phenomena and by IR excesses  from circumstellar disk dust.
Even if our sample is dominated by WTTS, we find that some objects show evident blue excesses with respect to
the locus of (V-I) vs. (B-V) colors, while eight YSOs have observed colors bluer  than the intrinsic ones
that would imply an unphysical negative reddening. This means
that  these objects are likely accretors in which the V-I colors include a contribution originating in the 
accretion hot
spot. For these objects we did not attempt to derive stellar parameters since we cannot estimate
the true interstellar absorption and get a reliable bolometric luminosity. 

By considering only the 78 YSOs with E(V-I)$>0$, we find a mean cluster reddening E(V-I)=0.7$\pm$0.5.
In addition, by comparing the ratios E(V-I)/E(B-V) and E(V-K)/E(B-V), we conclude that 
  the NGC\,6530 members  show an anomalous extinction with R$_{\rm} \lesssim$5.0. 
Therefore, for our analysis we used
 the reddening law given by \citet{muna96} for R$_{\rm V}$=5.0. The mean cluster reddening corresponds
to E(B-V)=0.42, which is consistent with the value
$<$E(B-V)$>$=0.35 found by \citet{sung00}. 
For this sample  we derived fundamental stellar parameters, i.e. effective temperatures, interstellar absorption, luminosities, ages, and masses. The HR diagram shows that our targets have ages between about 1 and 6-7\,Myr.
There is not a significant difference of ages between WTTS and CTTS, but if we distinguish the populations younger and
older than 2\,Myr, we find that the youngest population is mainly concentrated in the SE region, while the
oldest population is loosely clustered in the N and SW  regions. Since the ages
of two populations
are significantly different, we can conclude that there are two generations of stars in this region, formed
during different events. 
The SE region is therefore the most recent site of  star formation in the FOV studied in this work.
Our results confirm the scenario of sequential star formation already suggested in previous works.

\begin{acknowledgements}
 Part of this work was financially supported by the ASI contract I/044/10/0.
We wish to thank C. Argiroffi for her help in some steps of the spectral classification.
\end{acknowledgements}
 \bibliographystyle{aa}
  \bibliography{/Users/prisinzano/BIBLIOGRAPHY/bibdesk}

\begin{thebibliography}{42}
\expandafter\ifx\csname natexlab\endcsname\relax\def\natexlab#1{#1}\fi

\bibitem[{{Arias} {et~al.}(2006){Arias}, {Barb{\'a}}, {Ma{\'{\i}}z
  Apell{\'a}niz}, {Morrell}, \& {Rubio}}]{aria06}
{Arias}, J.~I., {Barb{\'a}}, R.~H., {Ma{\'{\i}}z Apell{\'a}niz}, J., {Morrell},
  N.~I., \& {Rubio}, M. 2006, \mnras, 366, 739

\bibitem[{{Arias} {et~al.}(2007){Arias}, {Barb{\'a}}, \& {Morrell}}]{aria07}
{Arias}, J.~I., {Barb{\'a}}, R.~H., \& {Morrell}, N.~I. 2007, \mnras, 374, 1253

\bibitem[{{Ballesteros-Paredes} {et~al.}(2007){Ballesteros-Paredes}, {Klessen},
  {Mac Low}, \& {Vazquez-Semadeni}}]{ball07}
{Ballesteros-Paredes}, J., {Klessen}, R.~S., {Mac Low}, M.-M., \&
  {Vazquez-Semadeni}, E. 2007, Protostars and Planets V, 63

\bibitem[{{Bessell}(1976)}]{bess76}
{Bessell}, M.~S. 1976, \pasp, 88, 557

\bibitem[{{Calvet} \& {Gullbring}(1998)}]{calv98}
{Calvet}, N. \& {Gullbring}, E. 1998, \apj, 509, 802

\bibitem[{{Conti} {et~al.}(1995){Conti}, {Hanson}, {Morris}, {Willis}, \&
  {Fossey}}]{cont95}
{Conti}, P.~S., {Hanson}, M.~M., {Morris}, P.~W., {Willis}, A.~J., \& {Fossey},
  S.~J. 1995, \apjl, 445, L35

\bibitem[{{Covino} {et~al.}(1997){Covino}, {Alcala}, {Allain}, {Bouvier},
  {Terranegra}, \& {Krautter}}]{covi97}
{Covino}, E., {Alcala}, J.~M., {Allain}, S., {et~al.} 1997, \aap, 328, 187

\bibitem[{{Da Rio} {et~al.}(2010){Da Rio}, {Robberto}, {Soderblom}, {Panagia},
  {Hillenbrand}, {Palla}, \& {Stassun}}]{dari10}
{Da Rio}, N., {Robberto}, M., {Soderblom}, D.~R., {et~al.} 2010, \apj, 722,
  1092

\bibitem[{{Damiani} {et~al.}(2004){Damiani}, {Flaccomio}, {Micela},
  {Sciortino}, {Harnden}, \& {Murray}}]{dami04}
{Damiani}, F., {Flaccomio}, E., {Micela}, G., {et~al.} 2004, \apj, 608, 781

\bibitem[{{Damiani} {et~al.}(2006){Damiani}, {Prisinzano}, {Micela}, \&
  {Sciortino}}]{dami06}
{Damiani}, F., {Prisinzano}, L., {Micela}, G., \& {Sciortino}, S. 2006, \aap,
  459, 477

\bibitem[{{Edvardsson}(1988)}]{edva88}
{Edvardsson}, B. 1988, \aap, 190, 148

\bibitem[{{Elmegreen}(2000)}]{elme00}
{Elmegreen}, B.~G. 2000, \apj, 530, 277

\bibitem[{{Elmegreen}(2007)}]{elme07}
{Elmegreen}, B.~G. 2007, \apj, 668, 1064

\bibitem[{{Fernandes} {et~al.}(2012){Fernandes}, {Gregorio-Hetem}, \&
  {Hetem}}]{fern12}
{Fernandes}, B., {Gregorio-Hetem}, J., \& {Hetem}, Jr, A. 2012, ArXiv e-prints

\bibitem[{{Fiorucci} \& {Munari}(2003)}]{fior03}
{Fiorucci}, M. \& {Munari}, U. 2003, \aap, 401, 781

\bibitem[{{Fitzpatrick}(1999)}]{fitz99}
{Fitzpatrick}, E.~L. 1999, \pasp, 111, 63

\bibitem[{{Fitzpatrick} \& {Massa}(2009)}]{fitz09}
{Fitzpatrick}, E.~L. \& {Massa}, D. 2009, \apj, 699, 1209

\bibitem[{{Gullbring} {et~al.}(2000){Gullbring}, {Calvet}, {Muzerolle}, \&
  {Hartmann}}]{gull00}
{Gullbring}, E., {Calvet}, N., {Muzerolle}, J., \& {Hartmann}, L. 2000, \apj,
  544, 927

\bibitem[{{Hartmann} {et~al.}(2001){Hartmann}, {Ballesteros-Paredes}, \&
  {Bergin}}]{hart01b}
{Hartmann}, L., {Ballesteros-Paredes}, J., \& {Bergin}, E.~A. 2001, \apj, 562,
  852

\bibitem[{{Huff} \& {Stahler}(2007)}]{huff07}
{Huff}, E.~M. \& {Stahler}, S.~W. 2007, \apj, 666, 281

\bibitem[{{Jao} {et~al.}(2008){Jao}, {Henry}, {Beaulieu}, \&
  {Subasavage}}]{jao08}
{Jao}, W.-C., {Henry}, T.~J., {Beaulieu}, T.~D., \& {Subasavage}, J.~P. 2008,
  \aj, 136, 840

\bibitem[{{Kenyon} \& {Hartmann}(1995)}]{keny95}
{Kenyon}, S.~J. \& {Hartmann}, L. 1995, \apjs, 101, 117

\bibitem[{{Kumar} \& {Anandarao}(2010)}]{kuma10}
{Kumar}, D.~L. \& {Anandarao}, B.~G. 2010, \mnras, 407, 1170

\bibitem[{{Lada} {et~al.}(1976){Lada}, {Gottlieb}, {Gottlieb}, \&
  {Gull}}]{lada76}
{Lada}, C.~J., {Gottlieb}, C.~A., {Gottlieb}, E.~W., \& {Gull}, T.~R. 1976,
  \apj, 203, 159

\bibitem[{{Leggett}(1992)}]{legg92}
{Leggett}, S.~K. 1992, \apjs, 82, 351

\bibitem[{{Luhman}(1999)}]{luhm99}
{Luhman}, K.~L. 1999, \apj, 525, 466

\bibitem[{{Mathis}(1990)}]{math90}
{Mathis}, J.~S. 1990, \araa, 28, 37

\bibitem[{{Meyer} {et~al.}(1997){Meyer}, {Calvet}, \& {Hillenbrand}}]{meye97}
{Meyer}, M.~R., {Calvet}, N., \& {Hillenbrand}, L.~A. 1997, \aj, 114, 288

\bibitem[{{Molinari} {et~al.}(2010){Molinari}, {Swinyard}, {Bally}, {Barlow},
  {Bernard}, {Martin}, {Moore}, {Noriega-Crespo}, {Plume}, {Testi}, {Zavagno},
  {Abergel}, {Ali}, {Anderson}, {Andr{\'e}}, {Baluteau}, {Battersby},
  {Beltr{\'a}n}, {Benedettini}, {Billot}, {Blommaert}, {Bontemps}, {Boulanger},
  {Brand}, {Brunt}, {Burton}, {Calzoletti}, {Carey}, {Caselli}, {Cesaroni},
  {Cernicharo}, {Chakrabarti}, {Chrysostomou}, {Cohen}, {Compiegne}, {de
  Bernardis}, {de Gasperis}, {di Giorgio}, {Elia}, {Faustini}, {Flagey},
  {Fukui}, {Fuller}, {Ganga}, {Garcia-Lario}, {Glenn}, {Goldsmith}, {Griffin},
  {Hoare}, {Huang}, {Ikhenaode}, {Joblin}, {Joncas}, {Juvela}, {Kirk},
  {Lagache}, {Li}, {Lim}, {Lord}, {Marengo}, {Marshall}, {Masi}, {Massi},
  {Matsuura}, {Minier}, {Miville-Desch{\^e}nes}, {Montier}, {Morgan}, {Motte},
  {Mottram}, {M{\"u}ller}, {Natoli}, {Neves}, {Olmi}, {Paladini}, {Paradis},
  {Parsons}, {Peretto}, {Pestalozzi}, {Pezzuto}, {Piacentini}, {Piazzo},
  {Polychroni}, {Pomar{\`e}s}, {Popescu}, {Reach}, {Ristorcelli}, {Robitaille},
  {Robitaille}, {Rod{\'o}n}, {Roy}, {Royer}, {Russeil}, {Saraceno}, {Sauvage},
  {Schilke}, {Schisano}, {Schneider}, {Schuller}, {Schulz}, {Sibthorpe},
  {Smith}, {Smith}, {Spinoglio}, {Stamatellos}, {Strafella}, {Stringfellow},
  {Sturm}, {Taylor}, {Thompson}, {Traficante}, {Tuffs}, {Umana}, {Valenziano},
  {Vavrek}, {Veneziani}, {Viti}, {Waelkens}, {Ward-Thompson}, {White},
  {Wilcock}, {Wyrowski}, {Yorke}, \& {Zhang}}]{moli10}
{Molinari}, S., {Swinyard}, B., {Bally}, J., {et~al.} 2010, \aap, 518, L100

\bibitem[{{Munari} \& {Carraro}(1996)}]{muna96}
{Munari}, U. \& {Carraro}, G. 1996, \aap, 314, 108

\bibitem[{{Palla} \& {Stahler}(2000)}]{pall00}
{Palla}, F. \& {Stahler}, S.~W. 2000, \apj, 540, 255

\bibitem[{{Povich} {et~al.}(2011){Povich}, {Townsley}, {Broos}, {Gagn{\'e}},
  {Babler}, {Indebetouw}, {Majewski}, {Meade}, {Getman}, {Robitaille}, \&
  {Townsend}}]{povi11}
{Povich}, M.~S., {Townsley}, L.~K., {Broos}, P.~S., {et~al.} 2011, \apjs, 194,
  6

\bibitem[{{Preibisch}(2012)}]{prei12}
{Preibisch}, T. 2012, Research in Astronomy and Astrophysics, 12, 1

\bibitem[{{Prisinzano} {et~al.}(2007){Prisinzano}, {Damiani}, {Micela}, \&
  {Pillitteri}}]{pris07}
{Prisinzano}, L., {Damiani}, F., {Micela}, G., \& {Pillitteri}, I. 2007, \aap,
  462, 123

\bibitem[{{Prisinzano} {et~al.}(2005){Prisinzano}, {Damiani}, {Micela}, \&
  {Sciortino}}]{pris05}
{Prisinzano}, L., {Damiani}, F., {Micela}, G., \& {Sciortino}, S. 2005, \aap,
  430, 941

\bibitem[{{S{\'a}nchez-Bl{\'a}zquez} {et~al.}(2006){S{\'a}nchez-Bl{\'a}zquez},
  {Peletier}, {Jim{\'e}nez-Vicente}, {Cardiel}, {Cenarro},
  {Falc{\'o}n-Barroso}, {Gorgas}, {Selam}, \& {Vazdekis}}]{sanc06}
{S{\'a}nchez-Bl{\'a}zquez}, P., {Peletier}, R.~F., {Jim{\'e}nez-Vicente}, J.,
  {et~al.} 2006, \mnras, 371, 703

\bibitem[{{Siess} {et~al.}(2000){Siess}, {Dufour}, \& {Forestini}}]{sies00}
{Siess}, L., {Dufour}, E., \& {Forestini}, M. 2000, \aap, 358, 593

\bibitem[{{Soderblom}(2010)}]{sode10}
{Soderblom}, D.~R. 2010, \araa, 48, 581

\bibitem[{{Solomon} {et~al.}(1979){Solomon}, {Sanders}, \& {Scoville}}]{solo79}
{Solomon}, P.~M., {Sanders}, D.~B., \& {Scoville}, N.~Z. 1979, in IAU
  Symposium, Vol.~84, The Large-Scale Characteristics of the Galaxy, ed.
  {W.~B.~Burton}, 35--52

\bibitem[{{Sung} {et~al.}(2000){Sung}, {Chun}, \& {Bessell}}]{sung00}
{Sung}, H., {Chun}, M.-Y., \& {Bessell}, M.~S. 2000, \aj, 120, 333

\bibitem[{{Tan} {et~al.}(2006){Tan}, {Krumholz}, \& {McKee}}]{tan06}
{Tan}, J.~C., {Krumholz}, M.~R., \& {McKee}, C.~F. 2006, \apjl, 641, L121

\bibitem[{{Wallerstein} \& {Cardelli}(1987)}]{wall87}
{Wallerstein}, G. \& {Cardelli}, J.~A. 1987, \aj, 93, 1522

\end{thebibliography}
 \longtabL{1}{
\begin{landscape}
\begin{longtable}{ccccccccc}
\caption{\label{targettable} List of stars observed with VIMOS, where ID is the sequential number,
ID$_{\rm WFI}$ is the \citet{pris05} identification number;
B, V,  and I 
 are the magnitudes 
presented in \citet{pris05},
while J, H, and K are the 2MASS magnitudes.}\\
\hline\hline
ID & $\alpha$(2000)~~~~~~$\delta$(2000)  & ID$_{\rm WFI}$ & B & V & I &J & H &K \\
\hline 
\endfirsthead
\caption{continued.}\\
\hline\hline
ID & $\alpha$(2000)~~~~~~$\delta$(2000)  & ID$_{\rm WFI}$ & B & V & I &J & H &K \\
\hline
\endhead
\hline
\endfoot
   1     &    18 03 47.18  -24 24 21.1     & 16351     & 17.394 $\pm$  0.003     & 16.109 $\pm$  0.002     & 14.409 $\pm$  0.005     &  13.02 $\pm$   0.03     &  12.14 $\pm$   0.05     &  11.61 $\pm$   0.04    \\
   2     &    18 03 49.33  -24 23 37.3     & 17591     & 18.059 $\pm$  0.012     & 16.762 $\pm$  0.004     & 15.111 $\pm$  0.008     &  13.71 $\pm$   0.03     &  12.98 $\pm$   0.04     &  12.67 $\pm$   0.05    \\
   3     &    18 03 51.02  -24 27 51.2     & 12265     & 20.192 $\pm$  0.013     & 18.480 $\pm$  0.010     & 16.219 $\pm$  0.006     &  14.64 $\pm$   0.02     &  13.71 $\pm$   0.05     &  13.29 $\pm$   0.06    \\
   4     &    18 03 52.02  -24 22 33.5     & 19447     & 19.178 $\pm$  0.021     & 17.931 $\pm$  0.010     & 15.915 $\pm$  0.004     &  14.36 $\pm$   0.02     &  13.38 $\pm$   0.04     &  12.94 $\pm$   0.05    \\
   5     &    18 03 54.06  -24 23 57.1     & 16985     & 19.241 $\pm$  0.009     & 17.743 $\pm$  0.006     & 15.699 $\pm$  0.005     &  13.97 $\pm$   0.06     &    ...     &  12.73 $\pm$   0.10    \\
   6     &    18 03 55.11  -24 19 18.6     & 25249     & 20.231 $\pm$  0.028     & 18.576 $\pm$  0.008     & 16.331 $\pm$  0.005     &  14.81 $\pm$   0.03     &  13.88 $\pm$   0.07     &    ...    \\
   7     &    18 03 55.24  -24 16 59.9     & 28700     & 15.733 $\pm$  0.002     & 14.700 $\pm$  0.002     & 13.429 $\pm$  0.012     &  12.47 $\pm$   0.04     &  11.92 $\pm$   0.05     &  11.75 $\pm$   0.04    \\
   8     &    18 03 55.84  -24 18 38.7     & 26138     &    ...     & 18.877 $\pm$  0.080     & 16.303 $\pm$  0.005     &  14.47 $\pm$   0.14     &  14.01 $\pm$   0.06     &  13.86 $\pm$   0.09    \\
   9     &    18 03 57.28  -24 16 09.8     & 30517     & 17.468 $\pm$  0.003     & 16.317 $\pm$  0.002     & 14.871 $\pm$  0.003     &    ...     &  12.68 $\pm$   0.09     &    ...    \\
  10     &    18 03 57.99  -24 24 41.9     & 15815     & 17.491 $\pm$  0.006     & 16.143 $\pm$  0.002     & 14.465 $\pm$  0.007     &  12.59 $\pm$   0.03     &  11.75 $\pm$   0.05     &  11.39 $\pm$   0.04    \\
  11     &    18 03 58.95  -24 18 17.7     & 26577     & 18.903 $\pm$  0.019     & 17.480 $\pm$  0.019     & 15.398 $\pm$  0.004     &  14.02 $\pm$   0.05     &  13.00 $\pm$   0.06     &  12.41 $\pm$   0.04    \\
  12     &    18 04 00.35  -24 25 30.4     & 14682     & 18.246 $\pm$  0.006     & 16.795 $\pm$  0.002     & 14.840 $\pm$  0.010     &  13.39 $\pm$   0.04     &    ...     &    ...    \\
  13     &    18 04 02.20  -24 17 53.4     & 27121     & 20.272 $\pm$  0.050     & 18.735 $\pm$  0.007     & 16.509 $\pm$  0.004     &  15.06 $\pm$   0.05     &  14.24 $\pm$   0.06     &    ...    \\
  14     &    18 04 02.94  -24 26 44.1     & 13412     & 18.626 $\pm$  0.013     & 17.119 $\pm$  0.004     & 15.038 $\pm$  0.007     &  13.36 $\pm$   0.05     &  12.55 $\pm$   0.06     &  12.31 $\pm$   0.02    \\
  15     &    18 04 04.10  -24 13 33.0     & 36413     & 17.671 $\pm$  0.003     & 16.354 $\pm$  0.002     & 14.762 $\pm$  0.003     &    ...     &    ...     &    ...    \\
  16     &    18 04 04.28  -24 17 10.5     & 28352     & 19.551 $\pm$  0.007     & 17.871 $\pm$  0.003     & 15.571 $\pm$  0.005     &    ...     &  12.93 $\pm$   0.05     &    ...    \\
  17     &    18 04 04.87  -24 27 32.0     & 12576     & 19.413 $\pm$  0.011     & 17.852 $\pm$  0.004     & 15.973 $\pm$  0.004     &  14.56 $\pm$   0.03     &  13.77 $\pm$   0.05     &    ...    \\
  18     &    18 04 05.08  -24 16 41.9     & 29284     & 17.659 $\pm$  0.002     & 16.276 $\pm$  0.002     & 14.539 $\pm$  0.007     &  13.41 $\pm$   0.02     &  12.59 $\pm$   0.04     &  12.27 $\pm$   0.04    \\
  19     &    18 04 05.30  -24 18 29.5     & 26329     & 19.246 $\pm$  0.023     & 17.908 $\pm$  0.020     & 16.074 $\pm$  0.011     &  14.05 $\pm$   0.03     &  13.37 $\pm$   0.03     &  12.99 $\pm$   0.03    \\
  20     &    18 04 05.74  -24 14 54.8     & 33362     & 18.448 $\pm$  0.005     & 17.040 $\pm$  0.003     & 15.026 $\pm$  0.007     &  13.51 $\pm$   0.03     &  12.70 $\pm$   0.04     &  12.45 $\pm$   0.04    \\
  21     &    18 04 05.80  -24 25 18.2     & 14903     & 18.149 $\pm$  0.007     & 16.765 $\pm$  0.005     & 14.990 $\pm$  0.008     &  13.42 $\pm$   0.03     &  12.38 $\pm$   0.05     &  11.82 $\pm$   0.04    \\
  22     &    18 04 07.35  -24 22 21.7     & 19764     & 15.838 $\pm$  0.009     & 14.563 $\pm$  0.004     & 12.935 $\pm$  0.004     &  11.73 $\pm$   0.03     &  10.99 $\pm$   0.04     &    ...    \\
  23     &    18 04 08.11  -24 13 46.6     & 35898     & 17.707 $\pm$  0.003     & 15.844 $\pm$  0.003     & 13.612 $\pm$  0.005     &  11.67 $\pm$   0.03     &    ...     &  10.00 $\pm$   0.05    \\
  24     &    18 04 08.15  -24 26 04.8     & 14096     & 15.745 $\pm$  0.010     & 14.469 $\pm$  0.002     & 12.749 $\pm$  0.008     &  11.45 $\pm$   0.03     &  10.79 $\pm$   0.03     &  10.54 $\pm$   0.03    \\
  25     &    18 04 09.16  -24 27 20.1     & 12779     & 19.382 $\pm$  0.013     & 17.804 $\pm$  0.007     & 15.582 $\pm$  0.004     &    ...     &    ...     &    ...    \\
  26     &    18 04 09.30  -24 16 30.7     & 29683     & 19.801 $\pm$  0.010     & 18.206 $\pm$  0.007     & 16.082 $\pm$  0.003     &    ...     &    ...     &    ...    \\
  27     &    18 04 09.63  -24 27 09.8     & 12983     & 18.236 $\pm$  0.013     & 16.737 $\pm$  0.005     & 14.796 $\pm$  0.004     &    ...     &    ...     &    ...    \\
  28     &    18 04 09.76  -24 16 22.1     & 30020     & 20.290 $\pm$  0.026     & 18.602 $\pm$  0.008     & 16.337 $\pm$  0.003     &  14.73 $\pm$   0.06     &    ...     &    ...    \\
  29     &    18 04 09.98  -24 19 29.8     & 24932     & 18.389 $\pm$  0.006     & 16.800 $\pm$  0.003     & 14.748 $\pm$  0.003     &  13.60 $\pm$   0.04     &  12.40 $\pm$   0.05     &  11.54 $\pm$   0.03    \\
  30     &    18 04 10.06  -24 26 34.6     & 13568     & 16.944 $\pm$  0.004     & 15.764 $\pm$  0.006     & 14.276 $\pm$  0.003     &  13.23 $\pm$   0.02     &  12.55 $\pm$   0.03     &  12.40 $\pm$   0.04    \\
  31     &    18 04 10.52  -24 26 55.6     & 13212     & 17.115 $\pm$  0.009     & 15.689 $\pm$  0.009     & 13.676 $\pm$  0.011     &  12.09 $\pm$   0.04     &  11.10 $\pm$   0.05     &  10.48 $\pm$   0.04    \\
  32     &    18 04 12.06  -24 19 09.7     & 25505     & 19.165 $\pm$  0.009     & 17.495 $\pm$  0.005     & 15.275 $\pm$  0.005     &  13.66 $\pm$   0.03     &  12.77 $\pm$   0.04     &  12.56 $\pm$   0.06    \\
  33     &    18 04 12.47  -24 19 42.9     & 24550     & 17.081 $\pm$  0.008     & 15.773 $\pm$  0.006     & 14.076 $\pm$  0.003     &  12.83 $\pm$   0.02     &  12.09 $\pm$   0.02     &  11.95 $\pm$   0.02    \\
  34     &    18 04 13.04  -24 23 17.1     & 18235     & 18.980 $\pm$  0.009     & 17.471 $\pm$  0.003     & 15.547 $\pm$  0.004     &  14.03 $\pm$   0.03     &  13.13 $\pm$   0.02     &  12.77 $\pm$   0.03    \\
  35     &    18 04 13.05  -24 22 42.7     & 19208     & 18.667 $\pm$  0.004     & 17.340 $\pm$  0.005     & 15.601 $\pm$  0.004     &  14.38 $\pm$   0.04     &  13.62 $\pm$   0.07     &    ...    \\
  36     &    18 04 13.07  -24 26 13.2     & 13933     & 19.207 $\pm$  0.005     & 17.467 $\pm$  0.004     & 15.119 $\pm$  0.003     &  13.37 $\pm$   0.03     &  12.38 $\pm$   0.03     &  11.94 $\pm$   0.03    \\
  37     &    18 04 13.24  -24 13 58.6     & 35460     & 19.630 $\pm$  0.012     & 17.987 $\pm$  0.008     & 15.652 $\pm$  0.004     &  14.07 $\pm$   0.06     &  13.20 $\pm$   0.06     &  12.90 $\pm$   0.06    \\
  38     &    18 04 13.35  -24 24 50.9     & 15585     & 16.686 $\pm$  0.006     & 15.394 $\pm$  0.004     & 13.684 $\pm$  0.009     &  12.46 $\pm$   0.02     &  11.72 $\pm$   0.02     &  11.54 $\pm$   0.02    \\
  39     &    18 04 13.38  -24 15 13.5     & 32646     & 18.545 $\pm$  0.004     & 16.995 $\pm$  0.005     & 14.905 $\pm$  0.002     &    ...     &  12.50 $\pm$   0.11     &  12.25 $\pm$   0.06    \\
  40     &    18 04 13.75  -24 17 26.9     & 27870     & 19.540 $\pm$  0.020     & 17.882 $\pm$  0.008     & 15.561 $\pm$  0.006     &    ...     &    ...     &    ...    \\
  41     &    18 04 13.91  -24 26 23.3     & 13782     & 17.936 $\pm$  0.006     & 16.422 $\pm$  0.005     & 14.320 $\pm$  0.002     &  12.66 $\pm$   0.02     &  11.77 $\pm$   0.03     &  11.33 $\pm$   0.02    \\
  42     &    18 04 14.52  -24 15 31.9     & 31912     & 19.384 $\pm$  0.010     & 17.833 $\pm$  0.004     & 15.819 $\pm$  0.006     &    ...     &    ...     &    ...    \\
  43     &    18 04 14.84  -24 17 37.6     & 27558     & 19.919 $\pm$  0.020     & 18.243 $\pm$  0.007     & 15.937 $\pm$  0.004     &  14.40 $\pm$   0.05     &  13.57 $\pm$   0.06     &  13.32 $\pm$   0.06    \\
  44     &    18 04 14.84  -24 22 55.4     & 18825     & 20.322 $\pm$  0.019     & 18.526 $\pm$  0.007     & 16.205 $\pm$  0.004     &  14.49 $\pm$   0.06     &    ...     &  13.37 $\pm$   0.11    \\
  45     &    18 04 26.29  -24 28 48.6     & 11173     & 20.225 $\pm$  0.013     & 18.809 $\pm$  0.011     & 16.370 $\pm$  0.004     &  14.30 $\pm$   0.02     &  13.30 $\pm$   0.03     &  12.73 $\pm$   0.04    \\
  46     &    18 04 26.52  -24 19 04.1     & 25566     & 19.254 $\pm$  0.014     & 17.691 $\pm$  0.005     & 15.455 $\pm$  0.009     &  13.85 $\pm$   0.04     &  12.93 $\pm$   0.04     &  12.51 $\pm$   0.05    \\
  47     &    18 04 27.38  -24 22 12.7     & 20043     & 19.330 $\pm$  0.025     & 17.605 $\pm$  0.005     & 15.272 $\pm$  0.002     &  13.33 $\pm$   0.04     &  12.38 $\pm$   0.04     &  12.05 $\pm$   0.03    \\
  48     &    18 04 27.71  -24 24 31.8     & 16096     & 18.520 $\pm$  0.003     & 16.982 $\pm$  0.006     & 14.860 $\pm$  0.002     &  13.40 $\pm$   0.06     &  12.51 $\pm$   0.05     &  12.25 $\pm$   0.04    \\
  49     &    18 04 27.93  -24 27 59.6     & 12099     & 20.923 $\pm$  0.039     & 18.978 $\pm$  0.018     & 15.841 $\pm$  0.003     &  13.99 $\pm$   0.03     &  12.89 $\pm$   0.03     &  12.29 $\pm$   0.03    \\
  50     &    18 04 28.22  -24 25 37.5     & 14558     & 19.513 $\pm$  0.011     & 17.922 $\pm$  0.010     & 15.618 $\pm$  0.003     &    ...     &  13.15 $\pm$   0.09     &  12.85 $\pm$   0.04    \\
  51     &    18 04 29.15  -24 13 43.2     & 36020     & 20.700 $\pm$  0.022     & 18.987 $\pm$  0.010     & 16.821 $\pm$  0.004     &  14.57 $\pm$   0.07     &  13.74 $\pm$   0.07     &  13.29 $\pm$   0.05    \\
  52     &    18 04 29.32  -24 25 28.0     & 14728     & 16.704 $\pm$  0.108     & 15.550 $\pm$  0.031     & 13.868 $\pm$  0.003     &  12.50 $\pm$   0.04     &  11.77 $\pm$   0.05     &  11.53 $\pm$   0.04    \\
  53     &    18 04 29.88  -24 14 29.3     & 34294     & 19.197 $\pm$  0.005     & 17.812 $\pm$  0.009     & 15.508 $\pm$  0.006     &  13.80 $\pm$   0.05     &  12.67 $\pm$   0.05     &  12.04 $\pm$   0.04    \\
  54     &    18 04 31.72  -24 14 54.1     & 33389     & 19.877 $\pm$  0.013     & 18.121 $\pm$  0.005     & 15.597 $\pm$  0.005     &  13.96 $\pm$   0.04     &  13.08 $\pm$   0.04     &    ...    \\
  55     &    18 04 31.88  -24 24 40.1     & 15882     & 17.651 $\pm$  0.002     & 16.249 $\pm$  0.003     & 14.417 $\pm$  0.002     &  13.13 $\pm$   0.03     &  12.31 $\pm$   0.05     &  12.09 $\pm$   0.03    \\
  56     &    18 04 32.25  -24 24 10.0     & 16628     & 15.791 $\pm$  0.001     & 14.639 $\pm$  0.002     & 13.133 $\pm$  0.004     &  11.92 $\pm$   0.06     &  11.24 $\pm$   0.05     &  11.06 $\pm$   0.04    \\
  57     &    18 04 32.84  -24 16 22.0     & 30029     & 20.457 $\pm$  0.039     & 18.925 $\pm$  0.017     & 16.688 $\pm$  0.010     &  15.14 $\pm$   0.04     &  14.38 $\pm$   0.08     &    ...    \\
  58     &    18 04 33.54  -24 26 53.4     & 13251     & 18.952 $\pm$  0.008     & 17.390 $\pm$  0.009     & 15.236 $\pm$  0.003     &  13.44 $\pm$   0.05     &    ...     &    ...    \\
  59     &    18 04 33.60  -24 23 20.2     & 18132     & 20.339 $\pm$  0.117     & 18.679 $\pm$  0.030     & 16.176 $\pm$  0.009     &    ...     &  13.43 $\pm$   0.12     &  13.02 $\pm$   0.11    \\
  60     &    18 04 33.96  -24 18 27.3     & 26358     & 19.829 $\pm$  0.045     & 18.450 $\pm$  0.007     & 16.216 $\pm$  0.011     &  14.31 $\pm$   0.06     &    ...     &    ...    \\
  61     &    18 04 35.27  -24 17 42.5     & 27449     & 16.109 $\pm$  0.002     & 14.903 $\pm$  0.002     & 13.410 $\pm$  0.010     &  12.10 $\pm$   0.05     &  11.30 $\pm$   0.11     &  10.14 $\pm$   0.08    \\
  62     &    18 04 35.32  -24 27 30.1     & 12608     & 18.220 $\pm$  0.006     & 16.647 $\pm$  0.002     & 14.556 $\pm$  0.005     &  12.94 $\pm$   0.04     &  12.06 $\pm$   0.07     &  11.74 $\pm$   0.06    \\
  63     &    18 04 35.37  -24 15 48.9     & 31269     & 19.398 $\pm$  0.025     & 17.926 $\pm$  0.005     & 15.938 $\pm$  0.003     &    ...     &  13.41 $\pm$   0.10     &    ...    \\
  64     &    18 04 35.73  -24 26 40.4     & 58855     &    ...     & 16.283 $\pm$  0.120     & 14.726 $\pm$  0.009     &    ...     &    ...     &    ...    \\
  65     &    18 04 37.85  -24 19 19.8     & 25221     & 20.066 $\pm$  0.011     & 18.421 $\pm$  0.016     & 16.031 $\pm$  0.004     &    ...     &  13.16 $\pm$   0.13     &  12.86 $\pm$   0.08    \\
  66     &    18 04 38.93  -24 16 10.5     & 30485     & 18.955 $\pm$  0.004     & 17.567 $\pm$  0.006     & 15.669 $\pm$  0.003     &  14.05 $\pm$   0.04     &  13.28 $\pm$   0.05     &  13.08 $\pm$   0.06    \\
  67     &    18 04 39.16  -24 23 53.6     & 17095     & 18.424 $\pm$  0.008     & 17.257 $\pm$  0.006     & 15.690 $\pm$  0.013     &  14.51 $\pm$   0.06     &    NaN $\pm$   0.08     &    ...    \\
  68     &    18 04 39.31  -24 22 34.1     & 19423     & 19.203 $\pm$  0.007     & 17.619 $\pm$  0.005     & 15.440 $\pm$  0.005     &  13.95 $\pm$   0.04     &  13.11 $\pm$   0.05     &  12.88 $\pm$   0.06    \\
  69     &    18 04 39.42  -24 27 08.8     & 13000     & 19.385 $\pm$  0.005     & 17.641 $\pm$  0.006     & 15.223 $\pm$  0.002     &  13.35 $\pm$   0.07     &  12.26 $\pm$   0.06     &  11.72 $\pm$   0.03    \\
  70     &    18 04 40.60  -24 25 49.5     & 14346     & 17.555 $\pm$  0.003     & 16.403 $\pm$  0.004     & 14.825 $\pm$  0.003     &  13.50 $\pm$   0.04     &  12.87 $\pm$   0.06     &  12.40 $\pm$   0.08    \\
  71     &    18 04 40.63  -24 27 50.3     & 58080     & 18.072 $\pm$  0.009     & 16.572 $\pm$  0.002     & 14.592 $\pm$  0.003     &    ...     &    ...     &    ...    \\
  72     &    18 04 40.86  -24 19 59.9     & 24025     & 19.961 $\pm$  0.011     & 18.268 $\pm$  0.007     & 15.827 $\pm$  0.004     &  14.14 $\pm$   0.03     &  13.24 $\pm$   0.03     &  13.05 $\pm$   0.06    \\
  73     &    18 04 41.46  -24 18 54.4     & 25725     & 15.686 $\pm$  0.003     & 14.672 $\pm$  0.002     & 13.389 $\pm$  0.008     &  12.42 $\pm$   0.03     &  11.81 $\pm$   0.04     &  11.70 $\pm$   0.05    \\
  74     &    18 04 41.51  -24 23 39.7     & 17514     & 20.327 $\pm$  0.015     & 18.991 $\pm$  0.008     & 16.039 $\pm$  0.003     &  13.88 $\pm$   0.04     &  12.78 $\pm$   0.06     &  12.15 $\pm$   0.05    \\
  75     &    18 04 41.89  -24 26 18.2     & 13854     & 18.682 $\pm$  0.005     & 16.924 $\pm$  0.002     & 14.566 $\pm$  0.003     &  12.77 $\pm$   0.04     &    ...     &    ...    \\
  76     &    18 04 42.21  -24 22 47.4     & 19069     & 19.601 $\pm$  0.011     & 17.955 $\pm$  0.004     & 15.610 $\pm$  0.004     &  13.98 $\pm$   0.04     &  13.10 $\pm$   0.05     &  12.74 $\pm$   0.07    \\
  77     &    18 04 43.49  -24 27 38.2     & 12467     & 18.077 $\pm$  0.005     & 16.265 $\pm$  0.003     & 13.938 $\pm$  0.005     &    ...     &  10.78 $\pm$   0.06     &    ...    \\
  78     &    18 04 44.10  -24 14 38.8     & 33939     & 17.458 $\pm$  0.003     & 16.197 $\pm$  0.002     & 14.594 $\pm$  0.004     &  13.30 $\pm$   0.02     &  12.59 $\pm$   0.04     &  12.39 $\pm$   0.04    \\
  79     &    18 04 44.14  -24 15 24.9     & 32188     & 18.892 $\pm$  0.008     & 17.151 $\pm$  0.004     & 14.802 $\pm$  0.005     &  12.92 $\pm$   0.03     &  12.01 $\pm$   0.04     &  11.65 $\pm$   0.03    \\
  80     &    18 04 44.21  -24 15 38.0     & 31670     & 19.097 $\pm$  0.008     & 17.588 $\pm$  0.005     & 15.561 $\pm$  0.004     &  13.89 $\pm$   0.02     &  12.95 $\pm$   0.02     &  12.54 $\pm$   0.04    \\
  81     &    18 04 44.66  -24 27 18.7     & 12798     & 18.024 $\pm$  0.005     & 16.534 $\pm$  0.003     & 14.675 $\pm$  0.006     &  13.28 $\pm$   0.04     &  12.41 $\pm$   0.05     &  12.23 $\pm$   0.04    \\
  82     &    18 04 45.01  -24 28 20.9     & 11664     & 20.416 $\pm$  0.016     & 18.587 $\pm$  0.007     & 15.958 $\pm$  0.007     &  13.98 $\pm$   0.03     &  13.03 $\pm$   0.02     &  12.76 $\pm$   0.03    \\
  83     &    18 04 45.23  -24 16 57.4     & 28774     & 18.610 $\pm$  0.028     & 17.109 $\pm$  0.004     & 15.211 $\pm$  0.005     &    ...     &    ...     &  12.80 $\pm$   0.05    \\
  84     &    18 04 46.33  -24 28 06.4     & 11957     & 20.248 $\pm$  0.012     & 18.507 $\pm$  0.006     & 16.087 $\pm$  0.005     &    ...     &    ...     &  13.12 $\pm$   0.06    \\
  85     &    18 04 46.37  -24 26 07.7     & 14025     & 19.047 $\pm$  0.005     & 17.368 $\pm$  0.004     & 15.199 $\pm$  0.007     &  13.67 $\pm$   0.02     &  12.54 $\pm$   0.03     &  12.01 $\pm$   0.02    \\
  86     &    18 04 47.42  -24 18 01.3     & 26948     & 12.968 $\pm$  0.008     & 12.640 $\pm$  0.020     & 12.122 $\pm$  0.012     &  11.59 $\pm$   0.02     &  11.31 $\pm$   0.03     &  11.22 $\pm$   0.02    \\
  87     &    18 04 47.83  -24 23 30.4     & 17811     & 16.977 $\pm$  0.001     & 15.680 $\pm$  0.002     & 13.928 $\pm$  0.006     &  12.57 $\pm$   0.02     &  11.81 $\pm$   0.02     &  11.66 $\pm$   0.02    \\
  88     &    18 04 48.53  -24 22 23.6     & 19704     & 19.279 $\pm$  0.067     & 17.822 $\pm$  0.003     & 15.934 $\pm$  0.004     &  14.58 $\pm$   0.04     &  13.79 $\pm$   0.05     &  13.62 $\pm$   0.05    \\
  89     &    18 04 50.53  -24 18 40.1     & 26104     & 20.004 $\pm$  0.011     & 18.340 $\pm$  0.006     & 16.128 $\pm$  0.005     &  14.41 $\pm$   0.09     &  13.52 $\pm$   0.09     &  13.16 $\pm$   0.05    \\
  90     &    18 04 51.58  -24 25 15.6     & 14959     & 18.654 $\pm$  0.004     & 17.002 $\pm$  0.003     & 14.726 $\pm$  0.004     &  12.98 $\pm$   0.02     &  12.02 $\pm$   0.02     &  11.55 $\pm$   0.02    \\
  91     &    18 04 53.47  -24 25 57.7     & 14192     & 20.528 $\pm$  0.015     & 18.314 $\pm$  0.006     & 15.133 $\pm$  0.004     &  12.51 $\pm$   0.03     &  11.28 $\pm$   0.03     &  10.63 $\pm$   0.02    \\
  92     &    18 04 54.49  -24 14 07.2     & 35120     & 19.375 $\pm$  0.012     & 17.723 $\pm$  0.007     & 15.528 $\pm$  0.002     &    ...     &    ...     &    ...    \\
  93     &    18 04 57.55  -24 24 46.5     & 15690     & 20.139 $\pm$  0.012     & 18.803 $\pm$  0.013     & 16.445 $\pm$  0.008     &  14.68 $\pm$   0.04     &  13.65 $\pm$   0.06     &  13.18 $\pm$   0.05    \\
  94     &    18 04 57.65  -24 24 56.4     & 15444     & 19.208 $\pm$  0.019     & 17.620 $\pm$  0.004     & 15.450 $\pm$  0.005     &  13.87 $\pm$   0.03     &  12.89 $\pm$   0.07     &  12.39 $\pm$   0.05    \\
\end{longtable}
\end{landscape}
}
  \longtab{3}{
\begin{longtable}{ccccccccccc}
\caption{\label{membershiptab} Results on membership, spectral types, and H$\alpha$ properties.}\\
\hline\hline
ID & NIR & Target & membership & membership & Spectral & H$_\alpha$ FWZI[$\AA$] & H$_\alpha$ FWZI[$\AA$] & Notes \\
   & class      & type &  PDM2007 & this work      & type     &  PDM2007       & this work &\\
\hline 
\endfirsthead
\caption{continued.}\\
\hline\hline
ID & NIR & Target & membership & membership & Spectral & H$_\alpha$ FWZI[$\AA$] & H$_\alpha$ FWZI[$\AA$] & Notes\\
   & class      & type &  PDM2007 & this work      & type     &  PDM2007       & this work & \\
\hline
\endhead
\hline
\endfoot
   1     &    II     &    2     &     M     &     1      &    K2    --    K3     V     &     8.3     &     9.5     &                        \\
   2     &   III     &    1     &     M     &     1      &    K2    --    K3     V     &     6.5     &     6.0     &                        \\
   3     &   III     &    1     &    --     &     1      &    M0    --    M1     V     &     ...     &     6.0     &                        \\
   4     &    II     &    2     &     M     &     1      &  K7.5    --  M0.5     V     &     ...     &    14.0     &          BVI excess    \\
   5     &   III     &    1     &     M     &     1      &    K4    --    K6     V     &     ...     &     6.0     &                        \\
   6     &   III     &    1     &    --     &     1      &  K4.5    --    K6     V     &     ...     &     6.0     &                        \\
   7     &   III     &    1     &     M     &     1      &    K4    --    K5     V     &    14.0     &     6.0     &                        \\
   8     &   III     &    1     &    --     &    -1      &    F7    --    F9     V     &     ...     &     6.0     &                        \\
   9     &   III     &    1     &     M     &     1      &    K7    --    M0     V     &     2.7     &     6.0     &                        \\
  10     &    II     &    2     &     M     &     1      &    K4    --    K5     V     &     ...     &    -4.0     &                        \\
  11     &    II     &    2     &     M     &     1      &  M0.5    --    M1     V     &     ...     &     6.0     &                        \\
  12     &   III     &    1     &     M     &     1      &  K3.5    --  K4.5     V     &     ...     &    11.0     &             veiling    \\
  13     &   III     &    1     &    --     &     1      &    M0    --    M1     V     &     ...     &    11.0     &                        \\
  14     &   III     &    1     &     M     &     1      &    K5    --    K6     V     &     8.6     &    13.0     &                        \\
  15     &           &    3     &    NM     &    -1      &    K5    --    K7     V     &     ...     &    -7.0     &                        \\
  16     &   III     &    1     &     M     &     1      &    G7    --    G9     V     &     ...     &     5.0     &                        \\
  17     &   III     &    1     &     M     &     1      &    K4    --    K6     V     &     ...     &    -6.5     &                        \\
  18     &   0/I     &    2     &     M     &     1      &  M0.5    --    M1     V     &     ...     &     7.0     &                        \\
  19     &           &    3     &    NM     &     1      &    M0    --    M1     V     &     ...     &    17.0     &                        \\
  20     &   III     &    1     &     M     &     1      &  K7.5    --  M0.5     V     &     ...     &    13.0     &                        \\
  21     &    II     &    2     &     M     &     1      &  K2.5    --  K3.5     V     &    12.5     &    15.0     &                        \\
  22     &   III     &    1     &     M     &     1      &    K1    --    K3     V     &     ...     &     7.0     &                        \\
  23     &    II     &    2     &    --     &    -1      &    K0    --    K1     V     &     ...     &    -6.0     &                        \\
  24     &   III     &    1     &    M?     &     1      &    F9    --    G1     V     &     ...     &     6.0     &                        \\
  25     &   III     &    1     &     M     &     1      &    K7    --    M0     V     &     ...     &    12.0     &                        \\
  26     &   III     &    1     &    --     &     1      &  M0.5    --    M1     V     &     ...     &     6.0     &                        \\
  27     &   III     &    1     &     M     &     1      &    K4    --    K5     V     &     ...     &     6.0     &                        \\
  28     &   III     &    1     &    --     &     1      &    K1    --    K2     V     &     ...     &    12.0     &                        \\
  29     &    II     &    2     &     M     &     1      &    M0    --    M1     V     &     ...     &     6.0     &                        \\
  30     &   III     &    1     &     M     &     1      &    K0    --    K2     V     &     ...     &     6.0     &                        \\
  31     &   0/I     &    2     &     M     &     1      &          --                 &    15.6     &    11.0     &      emission lines    \\
  32     &   III     &    1     &     M     &     1      &  M0.5    --  M1.5     V     &     ...     &     6.0     &                        \\
  33     &   III     &    1     &     M     &     1      &    K0    --    K2     V     &     ...     &    15.0     &                        \\
  34     &    II     &    2     &     M     &     1      &    K4    --    K6     V     &     ...     &    10.0     &                        \\
  35     &   III     &    1     &     M     &     1      &    K4    --    K6     V     &     ...     &    13.0     &                        \\
  36     &    II     &    2     &     M     &     1      &  K7.5    --  M0.5     V     &     ...     &     9.5     &             veiling    \\
  37     &   III     &    1     &     M     &     1      &    M0    --  M0.5     V     &     ...     &     9.0     &                        \\
  38     &   III     &    1     &     M     &     1      &    K2    --    K3     V     &     ...     &    -7.0     &                        \\
  39     &   III     &    1     &     M     &     1      &    G9    --    K1     V     &    10.7     &     7.0     &                        \\
  40     &   III     &    1     &     M     &     1      &    M0    --    M1     V     &     ...     &    13.5     &                        \\
  41     &    II     &    2     &     M     &     1      &    K3    --    K5     V     &    12.0     &    14.0     &                        \\
  42     &   III     &    1     &     M     &     1      &  K7.5    --    M0     V     &     ...     &    10.0     &                        \\
  43     &   III     &    1     &    --     &     1      &    M0    --    M1     V     &     ...     &    13.0     &                        \\
  44     &   III     &    1     &    --     &     1      &    K4    --    K6     V     &     ...     &     7.0     &                        \\
  45     &    II     &    2     &    --     &     1      &    M0    --  M0.5     V     &     ...     &     6.0     &          BVI excess    \\
  46     &   III     &    1     &     M     &     1      &          --                 &     ...     &   -70.0     &        high rotator    \\
  47     &   III     &    1     &    --     &     1      &    M0    --  M0.5     V     &     ...     &    10.0     &                        \\
  48     &   III     &    1     &     M     &     1      &    M0    --  M0.5     V     &     ...     &    10.0     &                        \\
  49     &   III     &    1     &    --     &     1      &    K0    --    K1     V     &     ...     &    22.0     &                        \\
  50     &   III     &    1     &     M     &     1      &    M0    --  M0.5     V     &     ...     &    11.0     &                        \\
  51     &           &    3     &    --     &    -1      &          --                 &     ...     &     6.0     &                        \\
  52     &   III     &    1     &     M     &     1      &    K2    --    K3     V     &    11.0     &    11.0     &                        \\
  53     &    II     &    2     &     M     &     1      &    M0    --    M1     V     &     ...     &     6.0     &          BVI excess    \\
  54     &   III     &    1     &     M     &     1      &    K2    --    K3     V     &     ...     &     5.0     &                        \\
  55     &   III     &    1     &    --     &     1      &    K2    --    K4     V     &     ...     &    -4.5     &                        \\
  56     &   III     &    1     &     M     &     1      &    G7    --    G8     V     &     ...     &     9.0     &                        \\
  57     &   III     &    1     &    --     &     1      &  K7.5    --  M0.5     V     &     ...     &    17.0     &                        \\
  58     &   III     &    1     &     M     &     1      &    G9    --    K1     V     &     5.4     &     6.0     &                        \\
  59     &   III     &    1     &    --     &     1      &  M0.5    --    M1     V     &     ...     &    18.0     &                        \\
  60     &   III     &    1     &    --     &     1      &    M0    --  M0.5     V     &     ...     &     8.0     &          BVI excess    \\
  61     &    II     &    2     &     M     &     1      &    K5    --    K6     V     &     ...     &     6.0     &                        \\
  62     &   III     &    1     &     M     &     1      &    M0    --  M0.5     V     &     ...     &    14.0     &                        \\
  63     &   III     &    1     &     M     &     1      &  K7.5    --  M0.5     V     &     ...     &    15.0     &                        \\
  64     &   III     &    1     &     M     &     1      &    M0    --  M0.5     V     &     3.4     &     8.0     &                        \\
  65     &   III     &    1     &    --     &     1      &    M0    --    M1     V     &     ...     &    17.0     &                        \\
  66     &   III     &    1     &     M     &     1      &    K4    --    K5     V     &     ...     &    12.0     &                        \\
  67     &           &    3     &     M     &     1      &  M0.5    --  M1.5     V     &     2.2     &     6.0     &                        \\
  68     &   III     &    1     &     M     &     1      &    K7    --    M0     V     &     ...     &    14.0     &                        \\
  69     &    II     &    2     &     M     &     1      &    F7    --    F8     V     &     ...     &   -24.0     &                        \\
  70     &           &    3     &    NM     &    -1      &    G4    --    G4     V     &    18.4     &   -13.0     &                        \\
  71     &   III     &    1     &     M     &     1      &    K7    --    M0     V     &    12.9     &     8.0     &                        \\
  72     &   III     &    1     &    --     &     1      &  K0.5    --  K2.5     V     &     ...     &     8.0     &                        \\
  73     &   III     &    1     &     M     &     1      &    G4    --    G6     V     &     ...     &     8.0     &                        \\
  74     &   III     &    1     &    --     &     1      &  M2.5    --  M3.5   III     &     ...     &    26.0     &          BVI excess    \\
  75     &   III     &    1     &     M     &     1      &    K2    --    K3     V     &     ...     &    17.0     &                        \\
  76     &   III     &    1     &     M     &     1      &    M0    --  M0.5     V     &     ...     &    15.0     &                        \\
  77     &    II     &    2     &     M     &     1      &    K2    --    K3     V     &     ...     &     6.0     &                        \\
  78     &   III     &    1     &     M     &     1      &    M0    --    M1     V     &     ...     &    17.0     &                        \\
  79     &   0/I     &    2     &     M     &     1      &    M1    --    M2     V     &     ...     &    12.0     &                        \\
  80     &    II     &    2     &     M     &     1      &    M0    --    M1     V     &    14.9     &    14.0     &                        \\
  81     &   III     &    1     &     M     &     1      &    M0    --  M0.5     V     &     ...     &     8.0     &                        \\
  82     &   III     &    1     &    --     &     1      &  M2.5    --  M3.5   III     &     ...     &    10.0     &                        \\
  83     &   III     &    1     &     M     &     1      &    M0    --    M1     V     &     2.4     &     5.0     &                        \\
  84     &   III     &    1     &    --     &     1      &  K4.5    --  K5.5     V     &     ...     &    10.0     &                        \\
  85     &    II     &    2     &     M     &     1      &    K7    --    M0     V     &    12.3     &     7.0     &                        \\
  86     &   III     &    1     &    --     &     1      &  K7.5    --  M0.5     V     &     ...     &    19.0     &             veiling    \\
  87     &   III     &    1     &     M     &     1      &  K1.5    --  K2.5     V     &     ...     &     5.0     &                        \\
  88     &   III     &    1     &     M     &     1      &    K2    --    K4     V     &     ...     &     6.0     &                        \\
  89     &   III     &    1     &    --     &     1      &    M4    --    M5   III     &     ...     &     4.0     &                        \\
  90     &    II     &    2     &     M     &     1      &  K3.5    --  K4.5     V     &    13.3     &     8.0     &                        \\
  91     &    II     &    2     &    --     &     1      &    A9    --    F1     V     &     ...     &    30.0     &                        \\
  92     &   III     &    1     &    M?     &    -1      &    K4    --    K5     V     &     ...     &     6.0     &                        \\
  93     &    II     &    2     &    --     &     1      &    M0    --    M1     V     &     ...     &    14.0     &          BVI excess    \\
  94     &    II     &    2     &     M     &     1      &  K7.5    --  M0.5     V     &     ...     &     8.0     &                        \\
  \end{longtable}
\tablefoot{ ID is the sequential number,
column 2 is the class assigned based on the NIR excesses; 
column 3 describes the target type where 1 indicates candidate WTTS, 2 indicates candidate CTTS and 3 indicates
candidates selected only photometrically;
column 4 indicates the membership assigned in PDM2007;
column 5 indicates the membership assigned in this work; 
column 6 gives the spectral types;
columns 7 and 8 gives the full width zero intensity (FWZI) of the H$_\alpha$ line
measured in PDM2007 and in this work, respectively;
column 9 gives some notes.}
  }
  \longtabL{5}{
\begin{landscape}
\begin{longtable}{cccccccccccccc}
\caption{\label{finaltable} Fundamental stellar parameters for the cluster members studied in this work
derived as described in Sect.\,\ref{hrsect}.}\\
\hline\hline
ID & A$_{\rm V}$& log(T$_{\rm eff}$) & log(T$_{\rm eff}$)$_{\rm min}$ & log(T$_{\rm eff}$)$_{\rm max}$ &
log(L$_{\rm bol}/L_\odot$) & log(L$_{\rm bol}/L_\odot$)$_{\rm min}$ &
 log(L$_{\rm bol}/L_\odot$)$_{\rm max}$ & 
 log(t) & log(t)$_{\rm min}$ & log(t)$_{\rm max}$ & M/M$_\odot$ & M$_{\rm min}$/M$_\odot$ & M$_{\rm max}$/M$_\odot$\\
      &  & [K]&[K] &[K] &  &  &   & [yrs] & [yrs] & [yrs] & & & \\
\hline 
\endfirsthead
\caption{continued.}\\
\hline\hline
ID & A$_{\rm V}$& log(T$_{\rm eff}$) & log(T$_{\rm eff}$)$_{\rm min}$ & log(T$_{\rm eff}$)$_{\rm max}$ &
log(L$_{\rm bol}/L_\odot$) & log(L$_{\rm bol}/L_\odot$)$_{\rm min}$ &
 log(L$_{\rm bol}/L_\odot$)$_{\rm max}$ & 
 log(t) & log(t)$_{\rm min}$ & log(t)$_{\rm max}$ & M/M$_\odot$ & M$_{\rm min}$/M$_\odot$ & M$_{\rm max}$/M$_\odot$\\
      &  & [K]&[K] &[K] &  &  &   & [yrs] & [yrs] & [yrs] & & & \\
\hline
\endhead
\hline
\endfoot
   1     & 2.0 $\pm$ 0.1     & 3.683   & 3.675   & 3.690     & 0.623   & 0.581   & 0.666     & 6.318     & 6.207   & 6.423   & 2.0     & 1.9   & 2.0  \\
   2     & 1.8 $\pm$ 0.1     & 3.683   & 3.675   & 3.690     & 0.303   & 0.260   & 0.347     & 6.690     & 6.591   & 6.779   & 1.6     & 1.6   & 1.6  \\
   3     & 1.1 $\pm$ 0.2     & 3.578   & 3.571   & 3.585     &-0.303   &-0.401   &-0.206     & 6.323     & 6.224   & 6.464   & 0.5     & 0.5   & 0.6  \\
   4     & 0.7 $\pm$ 0.3     & 3.585   & 3.578   & 3.597     &-0.318   &-0.428   &-0.209     & 6.436     & 6.281   & 6.660   & 0.6     & 0.5   & 0.7  \\
   5     & 2.1 $\pm$ 0.5     & 3.638   & 3.624   & 3.662     & 0.109   &-0.090   & 0.308     & 6.431     & 6.106   & 6.759   & 1.1     & 0.8   & 1.3  \\
   6     & 2.7 $\pm$ 0.3     & 3.638   & 3.624   & 3.650     & 0.018   &-0.118   & 0.154     & 6.571     & 6.334   & 6.794   & 1.1     & 0.9   & 1.2  \\
   7     & 0.0 $\pm$ 0.3     & 3.650   & 3.638   & 3.662     & 0.487   & 0.360   & 0.614     & 6.079     & 5.900   & 6.268   & 1.3     & 1.1   & 1.6  \\
   9     &-0.8 $\pm$ 0.3   & 3.597     & 3.585     & 3.609     &   ...     & ...     & ...     &   ...     & ...     & ...     &   ...     & ...     & ...    \\
  10     & 1.3 $\pm$ 0.3     & 3.650   & 3.638   & 3.662     & 0.400   & 0.273   & 0.527     & 6.182     & 6.004   & 6.382   & 1.3     & 1.1   & 1.5  \\
  11     & 0.5 $\pm$ 0.1     & 3.574   & 3.571   & 3.578     &-0.149   &-0.202   &-0.096     & 6.112     & 6.080   & 6.168   & 0.5     & 0.5   & 0.5  \\
  12     & 2.4 $\pm$ 0.2     & 3.662   & 3.650   & 3.668     & 0.566   & 0.480   & 0.651     & 6.130     & 5.995   & 6.318   & 1.6     & 1.4   & 1.8  \\
  13     & 1.0 $\pm$ 0.2     & 3.578   & 3.571   & 3.585     &-0.447   &-0.544   &-0.350     & 6.519     & 6.395   & 6.673   & 0.5     & 0.5   & 0.6  \\
  14     & 2.0 $\pm$ 0.2     & 3.631   & 3.624   & 3.638     & 0.351   & 0.278   & 0.424     & 6.016     & 5.919   & 6.129   & 1.0     & 0.9   & 1.1  \\
  16     & 4.5 $\pm$ 0.1     & 3.742   & 3.733   & 3.751     & 0.841   & 0.816   & 0.866     & 6.709     & 6.607   & 6.807   & 2.0     & 1.8   & 2.1  \\
  17     & 1.6 $\pm$ 0.5     & 3.638   & 3.624   & 3.662     &-0.134   &-0.333   & 0.065     & 6.808     & 6.451   & 7.082   & 1.1     & 0.9   & 1.1  \\
  18     &-0.4 $\pm$ 0.1   & 3.578     & 3.571     & 3.578     &   ...     & ...     & ...     &   ...     & ...     & ...     &   ...     & ...     & ...    \\
  19     &-0.1 $\pm$ 0.3   & 3.578     & 3.571     & 3.585     &   ...     & ...     & ...     &   ...     & ...     & ...     &   ...     & ...     & ...    \\
  20     & 0.6 $\pm$ 0.3     & 3.585   & 3.578   & 3.597     & 0.036   &-0.073   & 0.144     & 6.010     & 5.915   & 6.158   & 0.6     & 0.5   & 0.6  \\
  21     & 2.1 $\pm$ 0.1     & 3.675   & 3.668   & 3.683     & 0.425   & 0.381   & 0.469     & 6.462     & 6.363   & 6.563   & 1.7     & 1.6   & 1.7  \\
  22     & 1.9 $\pm$ 0.2     & 3.690   & 3.675   & 3.706     & 1.181   & 1.091   & 1.272     & 5.875     & 5.664   & 6.074   & 2.9     & 2.6   & 2.9  \\
  24     & 3.0 $\pm$ 0.1     & 3.780   & 3.774   & 3.786     & 1.595   & 1.575   & 1.616     & 6.306     & 6.272   & 6.325   & 3.0     & 2.9   & 3.1  \\
  25     & 1.6 $\pm$ 0.3     & 3.597   & 3.585   & 3.609     & 0.035   &-0.086   & 0.156     & 6.100     & 5.948   & 6.292   & 0.6     & 0.6   & 0.8  \\
  26     & 0.7 $\pm$ 0.1     & 3.578   & 3.571   & 3.578     &-0.359   &-0.408   &-0.309     & 6.397     & 6.330   & 6.470   & 0.5     & 0.5   & 0.6  \\
  27     & 2.1 $\pm$ 0.3     & 3.650   & 3.638   & 3.662     & 0.480   & 0.353   & 0.606     & 6.087     & 5.910   & 6.278   & 1.3     & 1.1   & 1.5  \\
  28     & 3.9 $\pm$ 0.1     & 3.698   & 3.690   & 3.706     & 0.371   & 0.322   & 0.420     & 6.768     & 6.667   & 6.864   & 1.6     & 1.6   & 1.7  \\
  29     & 0.5 $\pm$ 0.2     & 3.578   & 3.571   & 3.585     & 0.117   & 0.021   & 0.214     & 5.889     & 5.767   & 5.965   & 0.5     & 0.5   & 0.6  \\
  30     & 1.7 $\pm$ 0.2     & 3.706   & 3.690   & 3.720     & 0.609   & 0.512   & 0.705     & 6.599     & 6.392   & 6.813   & 1.9     & 1.7   & 2.0  \\
  32     & 0.8 $\pm$ 0.3     & 3.571   & 3.562   & 3.578     &-0.019   &-0.122   & 0.084     & 5.985     & 5.945   & 6.044   & 0.5     & 0.4   & 0.5  \\
  33     & 2.3 $\pm$ 0.2     & 3.706   & 3.690   & 3.720     & 0.857   & 0.760   & 0.953     & 6.330     & 6.122   & 6.561   & 2.2     & 2.1   & 2.2  \\
  34     & 1.7 $\pm$ 0.5     & 3.638   & 3.624   & 3.662     & 0.073   &-0.126   & 0.272     & 6.485     & 6.153   & 6.794   & 1.1     & 0.8   & 1.3  \\
  35     & 1.1 $\pm$ 0.5     & 3.638   & 3.624   & 3.662     &-0.098   &-0.296   & 0.101     & 6.750     & 6.395   & 7.036   & 1.1     & 0.9   & 1.1  \\
  36     & 1.7 $\pm$ 0.3     & 3.585   & 3.578   & 3.597     & 0.267   & 0.159   & 0.376     & 5.816     & 5.766   & 5.915   & 0.6     & 0.5   & 0.6  \\
  37     & 1.5 $\pm$ 0.1     & 3.582   & 3.578   & 3.585     & 0.013   &-0.036   & 0.063     & 6.004     & 5.958   & 6.051   & 0.5     & 0.5   & 0.6  \\
  38     & 2.0 $\pm$ 0.1     & 3.683   & 3.675   & 3.690     & 0.921   & 0.878   & 0.965     & 6.012     & 5.903   & 6.115   & 2.3     & 2.2   & 2.4  \\
  39     & 3.7 $\pm$ 0.2     & 3.720   & 3.706   & 3.733     & 0.914   & 0.853   & 0.974     & 6.526     & 6.357   & 6.723   & 2.2     & 2.0   & 2.4  \\
  40     & 1.3 $\pm$ 0.2     & 3.578   & 3.571   & 3.585     & 0.008   &-0.089   & 0.105     & 5.981     & 5.931   & 6.077   & 0.5     & 0.5   & 0.6  \\
  41     & 2.9 $\pm$ 0.4     & 3.662   & 3.638   & 3.675     & 0.892   & 0.723   & 1.061     & 5.789     & 5.588   & 6.156   & 1.8     & 1.4   & 2.2  \\
  42     & 0.6 $\pm$ 0.1     & 3.585   & 3.582   & 3.591     &-0.282   &-0.336   &-0.227     & 6.383     & 6.310   & 6.473   & 0.6     & 0.5   & 0.6  \\
  43     & 1.3 $\pm$ 0.2     & 3.578   & 3.571   & 3.585     &-0.154   &-0.251   &-0.057     & 6.148     & 6.069   & 6.266   & 0.5     & 0.5   & 0.6  \\
  44     & 2.9 $\pm$ 0.5     & 3.638   & 3.624   & 3.662     & 0.129   &-0.070   & 0.328     & 6.401     & 6.080   & 6.732   & 1.1     & 0.8   & 1.3  \\
  45     & 1.8 $\pm$ 0.1     & 3.582   & 3.578   & 3.585     &-0.190   &-0.240   &-0.140     & 6.229     & 6.164   & 6.292   & 0.5     & 0.5   & 0.6  \\
  47     & 1.5 $\pm$ 0.1     & 3.582   & 3.578   & 3.585     & 0.164   & 0.115   & 0.212     & 5.876     & 5.844   & 5.909   & 0.5     & 0.5   & 0.6  \\
  48     & 0.8 $\pm$ 0.1     & 3.582   & 3.578   & 3.585     & 0.159   & 0.110   & 0.208     & 5.881     & 5.847   & 5.914   & 0.5     & 0.5   & 0.6  \\
  49     & 6.8 $\pm$ 0.1     & 3.713   & 3.706   & 3.720     & 1.346   & 1.293   & 1.399     & 6.044     & 5.929   & 6.171   & 3.2     & 3.0   & 3.2  \\
  50     & 1.4 $\pm$ 0.1     & 3.582   & 3.578   & 3.585     & 0.002   &-0.048   & 0.052     & 6.017     & 5.969   & 6.064   & 0.5     & 0.5   & 0.6  \\
  52     & 1.9 $\pm$ 0.1     & 3.683   & 3.675   & 3.690     & 0.825   & 0.769   & 0.882     & 6.106     & 5.963   & 6.230   & 2.2     & 2.0   & 2.2  \\
  53     & 1.3 $\pm$ 0.2     & 3.578   & 3.571   & 3.585     & 0.016   &-0.081   & 0.113     & 5.975     & 5.924   & 6.069   & 0.5     & 0.5   & 0.6  \\
  54     & 4.5 $\pm$ 0.1     & 3.683   & 3.675   & 3.690     & 0.811   & 0.768   & 0.854     & 6.123     & 6.011   & 6.217   & 2.2     & 2.0   & 2.2  \\
  55     & 2.3 $\pm$ 0.2     & 3.675   & 3.662   & 3.690     & 0.700   & 0.616   & 0.785     & 6.140     & 5.955   & 6.332   & 1.9     & 1.7   & 2.0  \\
  56     & 2.1 $\pm$ 0.0     & 3.742   & 3.742   & 3.751     & 1.177   & 1.163   & 1.190     & 6.428     & 6.385   & 6.469   & 2.5     & 2.5   & 2.6  \\
  57     & 1.3 $\pm$ 0.3     & 3.585   & 3.578   & 3.597     &-0.450   &-0.561   &-0.339     & 6.622     & 6.449   & 6.878   & 0.6     & 0.5   & 0.7  \\
  58     & 3.9 $\pm$ 0.2     & 3.720   & 3.706   & 3.733     & 0.833   & 0.772   & 0.894     & 6.559     & 6.405   & 6.787   & 2.2     & 1.9   & 2.2  \\
  59     & 1.8 $\pm$ 0.2     & 3.574   & 3.571   & 3.578     &-0.121   &-0.183   &-0.060     & 6.084     & 6.052   & 6.145   & 0.5     & 0.5   & 0.5  \\
  60     & 1.2 $\pm$ 0.1     & 3.582   & 3.578   & 3.585     &-0.293   &-0.344   &-0.243     & 6.360     & 6.289   & 6.429   & 0.6     & 0.5   & 0.6  \\
  61     & 0.2 $\pm$ 0.2     & 3.631   & 3.624   & 3.638     & 0.529   & 0.455   & 0.602     & 5.829     & 5.744   & 5.918   & 1.0     & 0.9   & 1.1  \\
  62     & 0.8 $\pm$ 0.1     & 3.582   & 3.578   & 3.585     & 0.255   & 0.207   & 0.304     & 5.815     & 5.791   & 5.839   & 0.5     & 0.5   & 0.6  \\
  63     & 0.6 $\pm$ 0.3     & 3.585   & 3.578   & 3.597     &-0.350   &-0.459   &-0.241     & 6.478     & 6.320   & 6.709   & 0.6     & 0.5   & 0.7  \\
  64     &-0.9 $\pm$ 0.4   & 3.582     & 3.578     & 3.585     &   ...     & ...     & ...     &   ...     & ...     & ...     &   ...     & ...     & ...    \\
  65     & 1.5 $\pm$ 0.2     & 3.578   & 3.571   & 3.585     &-0.124   &-0.223   &-0.026     & 6.112     & 6.043   & 6.230   & 0.5     & 0.5   & 0.6  \\
  66     & 1.9 $\pm$ 0.3     & 3.650   & 3.638   & 3.662     & 0.096   &-0.031   & 0.222     & 6.616     & 6.400   & 6.816   & 1.3     & 1.1   & 1.3  \\
  67     &-1.2 $\pm$ 0.3   & 3.571     & 3.562     & 3.578     &   ...     & ...     & ...     &   ...     & ...     & ...     &   ...     & ...     & ...    \\
  68     & 1.4 $\pm$ 0.3     & 3.597   & 3.585   & 3.609     & 0.057   &-0.063   & 0.178     & 6.075     & 5.926   & 6.260   & 0.6     & 0.6   & 0.8  \\
  69     & 5.4 $\pm$ 0.1     & 3.795   & 3.792   & 3.798     & 1.244   & 1.218   & 1.269     & 6.708     & 6.692   & 6.724   & 2.1     & 2.1   & 2.2  \\
  71     & 0.8 $\pm$ 0.3     & 3.597   & 3.585   & 3.609     & 0.236   & 0.116   & 0.357     & 5.902     & 5.788   & 6.059   & 0.6     & 0.6   & 0.7  \\
  72     & 4.4 $\pm$ 0.2     & 3.698   & 3.683   & 3.713     & 0.717   & 0.623   & 0.811     & 6.391     & 6.176   & 6.563   & 2.1     & 2.0   & 2.1  \\
  73     & 1.6 $\pm$ 0.0     & 3.761   & 3.756   & 3.763     & 0.939   & 0.923   & 0.955     & 6.792     & 6.758   & 6.823   & 1.9     & 1.9   & 1.9  \\
  74     & 1.5 $\pm$ 0.5     & 3.540   & 3.534   & 3.547     &-0.110   &-0.327   & 0.107     & 5.578     & 5.289   & 5.867   & 0.4     & 0.3   & 0.4  \\
  75     & 4.0 $\pm$ 0.1     & 3.683   & 3.675   & 3.690     & 1.090   & 1.048   & 1.133     & 5.868     & 5.766   & 5.970   & 2.6     & 2.4   & 2.7  \\
  76     & 1.5 $\pm$ 0.1     & 3.582   & 3.578   & 3.585     & 0.038   &-0.010   & 0.087     & 5.983     & 5.938   & 6.027   & 0.5     & 0.5   & 0.6  \\
  77     & 3.9 $\pm$ 0.1     & 3.683   & 3.675   & 3.690     & 1.316   & 1.274   & 1.359     & 5.678     & 5.593   & 5.768   & 3.0     & 2.8   & 3.0  \\
  78     &-0.8 $\pm$ 0.2   & 3.578     & 3.571     & 3.585     &   ...     & ...     & ...     &   ...     & ...     & ...     &   ...     & ...     & ...    \\
  79     & 0.9 $\pm$ 0.3     & 3.562   & 3.554   & 3.571     & 0.208   & 0.099   & 0.316     & 5.560     & 5.292   & 5.743   & 0.4     & 0.4   & 0.5  \\
  80     & 0.4 $\pm$ 0.2     & 3.578   & 3.571   & 3.585     &-0.228   &-0.325   &-0.132     & 6.234     & 6.142   & 6.362   & 0.5     & 0.5   & 0.6  \\
  81     & 0.1 $\pm$ 0.1     & 3.582   & 3.578   & 3.585     & 0.021   &-0.028   & 0.070     & 5.998     & 5.952   & 6.043   & 0.5     & 0.5   & 0.6  \\
  82     & 0.5 $\pm$ 0.5     & 3.540   & 3.534   & 3.547     &-0.337   &-0.555   &-0.120     & 6.121     & 5.996   & 6.252   & 0.4     & 0.3   & 0.4  \\
  83     & 0.1 $\pm$ 0.2     & 3.578   & 3.571   & 3.585     &-0.192   &-0.289   &-0.095     & 6.189     & 6.107   & 6.313   & 0.5     & 0.5   & 0.6  \\
  84     & 3.2 $\pm$ 0.2     & 3.638   & 3.631   & 3.650     & 0.256   & 0.156   & 0.356     & 6.226     & 6.066   & 6.388   & 1.1     & 1.0   & 1.3  \\
  85     & 1.4 $\pm$ 0.3     & 3.597   & 3.585   & 3.609     & 0.146   & 0.025   & 0.267     & 5.982     & 5.857   & 6.154   & 0.6     & 0.6   & 0.7  \\
  86     &-3.9 $\pm$ 0.3   & 3.585     & 3.578     & 3.597     &   ...     & ...     & ...     &   ...     & ...     & ...     &   ...     & ...     & ...    \\
  87     & 2.2 $\pm$ 0.1     & 3.690   & 3.683   & 3.698     & 0.884   & 0.838   & 0.930     & 6.121     & 6.020   & 6.217   & 2.3     & 2.2   & 2.3  \\
  88     & 2.4 $\pm$ 0.2     & 3.675   & 3.662   & 3.690     & 0.139   & 0.054   & 0.223     & 6.830     & 6.671   & 6.996   & 1.4     & 1.3   & 1.4  \\
  89     &-2.8 $\pm$ 0.8   & 3.519     & 3.511     & 3.528     &   ...     & ...     & ...     &   ...     & ...     & ...     &   ...     & ...     & ...    \\
  90     & 3.4 $\pm$ 0.2     & 3.662   & 3.650   & 3.668     & 0.870   & 0.785   & 0.954     & 5.811     & 5.702   & 5.993   & 1.8     & 1.6   & 2.1  \\
  91     & 8.3 $\pm$ 0.1     & 3.857   & 3.848   & 3.869     & 2.131   & 2.088   & 2.174     & 6.011     & 5.975   & 6.034   & 4.0     & 3.9   & 4.1  \\
  93     & 1.4 $\pm$ 0.2     & 3.578   & 3.571   & 3.585     &-0.315   &-0.414   &-0.217     & 6.340     & 6.236   & 6.482   & 0.5     & 0.5   & 0.6  \\
  94     & 1.1 $\pm$ 0.3     & 3.585   & 3.578   & 3.597     &-0.008   &-0.117   & 0.100     & 6.054     & 5.952   & 6.216   & 0.6     & 0.5   & 0.6  \\
\end{longtable}
\end{landscape}
\tablefoot{ ID is the sequential number,
column 2 is the interstellar extinction; 
columns 3, 4, and 5 give the bolometric luminosities;
columns 6, 7, and 8 give the effective temperatures;
columns 9, 10, and 11 give the ages
and, finally, columns 12, 13, and 14 give masses. The relative uncertainty ranges are also given.}
}

\end{document}